\documentclass[a4paper,12pt]{article}

\ifx\pdfoutput\undefined
\usepackage[dvips,bookmarks=false]{hyperref}	
\else
\usepackage{hyperref}	
\fi
\hypersetup{colorlinks,bookmarksopen,bookmarksnumbered,citecolor=blue,
linkcolor=black,pdfstartview=FitH,urlcolor=blue}

\usepackage{graphicx}
\usepackage{amsmath}
\usepackage{amssymb}
\usepackage{cite}
\usepackage{braket}
\usepackage{bm}
\usepackage{color}
\usepackage{hhline}
\usepackage{multirow}
\usepackage{indentfirst}
\usepackage{comment}

\allowdisplaybreaks

\newcommand{\Slash}[1]{{\ooalign{\hfil/\hfil\crcr$#1$}}}
\newcommand{\tr}{\mathrm{tr}}

\newcommand{\1}{\mbox{1}\hspace{-0.25em}\mbox{l}}

\renewcommand{\cal}{\mathcal}

\renewcommand{\bar}{\overline}

\usepackage{ulem}

\begin{document}

\begin{titlepage}

\begin{flushright}
KUNS-2815
\end{flushright}

\begin{center}

\vspace{1cm}
{\large\textbf{
Implications of the weak gravity conjecture\\ in anomalous quiver gauge theories
}
 }
\vspace{1cm}

\renewcommand{\thefootnote}{\fnsymbol{footnote}}
Yoshihiko Abe$^{1}$\footnote[1]{y.abe@gauge.scphys.kyoto-u.ac.jp}
,
Tetsutaro Higaki$^{2}$\footnote[2]{thigaki@rk.phys.keio.ac.jp}
and Rei Takahashi$^{2}$
\vspace{5mm}

\textit{%
$^1${Department of Physics, Kyoto University, Kyoto 606-8502, Japan}\\
 $^2${Department of Physics, Keio University, Yokohama 223-8533, Japan}\\
}

\vspace{8mm}

\abstract{
We argue a smallness of gauge couplings in abelian quiver gauge theories, taking
the anomaly cancellation condition into account.
In theories of our interest there exist chiral fermions leading to chiral gauge anomalies, and
an anomaly-free gauge coupling
tends to be small, and hence can give a non-trivial condition of the weak gravity conjecture. 
As concrete examples, we consider $U(1)^{k}$ gauge theories 
with a discrete symmetry associated with cyclic permutations 
between the gauge groups, and identify anomaly-free $U(1)$ gauge symmetries and 
the corresponding gauge couplings.
Owing to this discrete symmetry, we can systematically study the models and we
find that the models would be examples of the weak coupling conjecture.
It is conjectured that a certain class of chiral gauge theories with too many $U(1)$ symmetries may be in the swampland.
We also numerically study constraints on the couplings from the scalar weak gravity conjecture in a concrete model.
These constraints may have a phenomenological implication to model building of a chiral hidden sector 
as well as the visible sector. 
}

\end{center}
\end{titlepage}

\renewcommand{\thefootnote}{\arabic{footnote}}
\newcommand{\bhline}[1]{\noalign{\hrule height #1}}
\newcommand{\bvline}[1]{\vrule width #1}

\setcounter{footnote}{0}

\setcounter{page}{1}

\section{Introduction}
\label{sec:1}

Swampland conjectures attract much attention recently in various apsects \cite{Vafa:2005ui,Ooguri:2006in,ArkaniHamed:2006dz,Obied:2018sgi,Garg:2018reu,Ooguri:2018wrx,Palti:2019pca}.
The conjectures are expected to constrain effective field theories
to be consistent with quantum gravity,
and give us new insights into not only the string theory as a candidate of quantum gravity
but also physics beyond the Standard Model (SM).
Among them, the weak gravity conjecture (WGC)
requires theories consistent with quantum gravity to include a charged state with 
a charge $q$ and a mass $m$ satisfying 
the weak gravity bound \cite{ArkaniHamed:2006dz},
\begin{align}
 e q \geq \frac{m}{\sqrt{2} M_{\mathrm{Pl}}},
\label{eq-WGC}
\end{align}
so that an extremal black holes can have a decay channel.
The WGC briefly states that the gravity is the weakest force.
Here, $e$ is an anomaly-free gauge coupling and $M_{\mathrm{Pl}}$ is the reduced Planck mass.
The WGC can be extended to theories with multiple $U(1)$ groups~\cite{Cheung:2014vva}
and also to 
a scalar exchange force such as a Yukawa interaction 
\cite{Palti:2017elp,Lust:2017wrl,Gonzalo:2019gjp,Andriot:2020lea,Benakli:2020pkm}.
The latter extension is called the scalar weak gravity conjecture (SWGC) .
These conjectures also have been checked in several aspects \cite{Lee:2018spm,Lee:2019tst},
and indicate that repulsive forces of gauge interactions among the same species of particles
are stronger than attractive forces of gravity and Yukawa interactions among them \cite{Heidenreich:2019zkl}.
The situation may not be so simple in chiral gauge theories.
IR symmetries are often obtained through the breaking of UV symmetries,
and an IR gauge coupling is given by a linear combination of UV gauge couplings as in the SM.
The linear combinations are determined by the St\"uckelberg couplings among 
the gauge bosons and would-be Nambu-Goldstone bosons (or axions) associated with the symmetry breaking.
This is applicable not only to anomaly-free gauge theories but
also to consistent theories possessing anomalous $U(1)$ gauge groups.
In theories with an anomalous $U(1)$, an axion field plays an important role to cancel the gauge anomalies: 
the gauge invariance is (non-linearly) restored owing to the axion coupling to topological terms of the gauge fields
on top of the St\"uckelberg couplings\footnote{
See also a recent work \cite{Craig:2019zkf}.
}.
As in the ordinary spontaneous symmetry breaking, 
these St\"uckelberg couplings lead to the gauge boson mass and 
determine the eigenstate of massless gauge boson.
Thus the gauge boson of anomalous $U(1)$ symmetry is decoupled in the low energy limit\footnote{
Some of gauge bosons in the anomaly-free gauge groups can also become massive through the St\"uckelberg couplings.
}.
In the string theory, this anomaly cancellation is realized by the Green-Schwarz mechanism \cite{Green:1984sg} 
involving string theoretic axions.
4D string models with anomalous $U(1)$'s have been well-discussed for realizing the SM \cite{Aldazabal:2000sa,Aldazabal:2000cn,Aldazabal:2000dg,Blumenhagen:2003jy,Blumenhagen:2005pm,Blumenhagen:2006ci}.
In this paper, we will focus on models with multiple $U(1)$ symmetries and chiral fermions.
For models with $U(1)^k$, an anomaly-free $U(1)$ is given by a linear combination of the original symmetries:
\begin{align}
 U(1)_{\text{anomaly-free}} = \sum_{i=1}^{k} c_{i} U(1)_{i},
\end{align}
where $k$ is the number of $U(1)$ symmetries, 
$c_{i}~(i=1,2, \ldots , k)$ is a model-dependent ${\cal O}(1)$ coefficient
and $U(1)_i$ is the $i$-th gauge group.
We will discuss some examples in the following section.
Then, the corresponding anomaly-free gauge coupling $e$ is given by
\begin{align}
 \frac{1}{e^2} = \sum_{i=1}^{k} \frac{c_{i}^2}{g_{i}^2},
 \label{eq:keyeq}
\end{align}
where $g_i$ is the gauge coupling of the $U(1)_i$ symmetry.
The gauge coupling $e$ will become necessarily very weak and smaller than the original coupling $g_i$ 
as the number of $U(1)$ gauge groups increases in the large $k$ limit\footnote{
The WGC with a similar gauge coupling is discussed in Ref.~\cite{Saraswat:2016eaz}.
}.
Thus, the WGC condition in Eq.~(\ref{eq-WGC}) looks hard to be satisfied
with an assumption that chiral anomlies can be canceled.
In other words, the repulsive force among particles will then become very weak.
It is conjectured that a certain class of chiral gauge theories with too many $U(1)$ symmetries 
can be in the swampland\footnote{
This will generally be applicable to theories with a semi-simple gauge group of $G=\prod_{i=1}^k G_i$
in the large $k$ limit, when $G$ is spontaneously broken to a simple group.
Here $G_i$ is a simple group. 
}.
It is noted that the gauge groups in 10D superstring theories are restricted \cite{Polchinski:1998rq},  
while those in 4D brane models seem less-constrained in the view point of tadpole condition\footnote{
In the heterotic string, the rank of the gauge group is sixteen.
}.
When magnitude of all the gauge couplings is comparable to each other,
the Eq.~(\ref{eq:keyeq}) is rewitten as
\begin{align}
 e q \sim \frac{\tilde{g}}{\sqrt{k}} q \gtrsim \frac{m}{M_{\mathrm{Pl}}},
 \label{eq:keyeq2}
\end{align}
where $\tilde{g} \sim g_i $ for $\forall~ i$ is the average of the gauge couplings. 
We find that the gauge coupling is scaling as 
$e \sim k^{-1/2}$ for a large $k$ and there exists an upper bound on $k$,
$k \lesssim (q{\tilde g}\frac{M_{\mathrm{Pl}}}{m})^2$, if the WGC is correct 
and the mass $m$ remains non-zero in the large $k$ limit.
This upper bound on $k$ is similar to the species bound \cite{Dvali:2007wp}, 
but $k$ is not the number of species but the number of $U(1)$ gauge groups in our case\footnote{
If we have too large $c_i$'s, the theory would be in the swampland
owing to the appearance of very weak coupling.}.
Similar conditions for theories with a discrete $\mathbb{Z}_{k}$ (gauge) symmetry 
are also discussed in Refs.~\cite{Craig:2018yvw,Buratti:2020kda}.
Eq.~(\ref{eq:keyeq2}) could be regarded as an example of the weak coupling conjecture \cite{Buratti:2020kda}.
A notion of quiver gauge theory is often used for theories in the presence of 
multiple gauge groups and bi-fundamental chiral fermions,
and matches model building involving D-branes well \cite{Ibanez:2001nd,Cremades:2002te,Cremades:2002cs,Verlinde:2005jr,Cvetic:2009yh,Krippendorf:2010hj,Dolan:2011qu,Douglas:1996sw,Uranga:2000ck,Burrington:2006uu,Yamazaki:2008bt}. 
Instead of concrete string models, in this paper we will
consider quiver gauge theories with $U(1)^{k}$ gauge groups and
focus on the anomaly-free gauge groups and the (S)WGC in a bottom-up approach, 
supposing that the remaining anomalies are canceled 
and then the anomalous gauge bosons get massive.
In general, computation of anomalies depends on the matter content in models.
In order to check anomaly-free $U(1)$'s systematically and study concretely the (S)WGC constraints on the gauge couplings, 
we restrict ourselves to several types of models controlled by
discrete symmetries.
However, a behavior of the anomaly-free gauge coupling in Eq.~(\ref{eq:keyeq}) does not change in general models with anomalous $U(1)$'s.
The (S)WGC can constrain range of free parameters in low energy theories 
and show what parameter values are favored by UV theory in the view point of IR physics.
In some quiver gauge theories of our interest, there exist a discrete symmetry associated with cyclic permutations 
between the gauge groups in certain quiver gauge theories,
and the symmetry can generally be broken in anomaly-free $U(1)$ theories
by a linear combination of $U(1)$'s as in Eq.~(\ref{eq:keyeq}).
Some of quiver gauge theories remind us of deconstructed extra dimension \cite{ArkaniHamed:2001ca,Abe:2002rj},
which could relate our approach to the weak coupling conjecture in holography \cite{Buratti:2020kda}. 
Also new insights can be given to chiral abelian gauge theories which may be a candidate of 
hidden sectors of dark matter models in particle physics \cite{Costa:2020dph}.
This paper is organized as follows.
In Sec.~2, 
we give a brief review of the (S)WGC and anomalous $U(1)$ symmetries.
In Sec.~3, 
we will discuss concrete quiver gauge theories with $U(1)^k$, then identify the anomaly-free $U(1)$ symmetries. 
In Sec.~4, 
we numerically show the SWGC constraint on the gauge couplings and Yukawa couplings in a $U(1)^4$ quiver gauge theory.
We discuss also a toy model from 5D orbifold compactification similarly.
Sec.~5 is devoted to summary and conclusion.
In this paper, we will discuss the above arguments with the tree level parameters. 
%


\section{Brief reviews of the (S)WGC and anomalous $U(1)$'s}

\subsection{The WGC and the SWGC}

In this subsection, we give a brief review of the WGC and the SWGC in four dimension.
The WGC claims 
that there exists a state with a charge $q$ and a mass $m$ 
satisfying the inequality
\begin{align}
 eq \geq \frac{m}{\sqrt{2} M_{\mathrm{Pl}}}
\end{align}
in a theory consistent with quantum gravity \cite{ArkaniHamed:2006dz}.
The factor of $1/\sqrt{2}$ comes from the relative normalization of the Newton force against the Coulomb one,
and a generalization to an arbitrary dimension is straightforward~\cite{Robinson:2006yd}.
This conjecture makes (super)extremal black holes decay into lighter ones.
The WGC can be extended to theories including a scalar exchange force such as a Yukawa interaction. 
This is called the SWGC \cite{Palti:2017elp,Lust:2017wrl,Lee:2018spm}.
Let us consider a theory with multiple $U(1)$ gauge groups:
\begin{align}
 S_{\mathrm{EM}} = \int d^4x\, \sqrt{-g} \biggl[
 \frac{M_{\mathrm{Pl}}^{2}}{2} \mathcal{R} 
 - \sum_{a,b} \frac{1}{2} K_{ab} \partial_{\mu} \phi^{a} \partial^{\mu} \phi^{b}
 -\frac{1}{4} \sum_{i,j} f_{ij}(\phi) F^{(i)}_{\mu\nu} F^{(j) \mu\nu} \biggr]
\end{align}
where $\mathcal{R}$ is a Ricci scalar, $\phi^a$ is a real scalar field, $F^{(i)}_{\mu \nu}$ is a field strength of $U(1)_i$,
$K_{ab}$ is a scalar kinetic matrix, $f_{ij}$ is a gauge kinetic function, and
$i,~j~(=1,2, \ldots, k)$ and $a,~b$ denote the labels of $U(1)$ gauge groups and those of scalar fields respectively.
The diagonal parts of $f_{ij}$ give the gauge couplings of $U(1)_{i}$'s and the off-diagonal components are kinetic mixings.
The matter part action is given by
\begin{align}
 S_{\mathrm{matter}} = \int d^4x\, \sqrt{-g}
 \biggl[
 -\frac{1}{2} \sum_{a, b} K_{ab} \partial_{\mu} \Phi^{a} \partial^{\mu} \Phi^{b}
 + \bar{\psi} i \gamma^{\mu} \Bigl( \nabla_{\mu} +i \sum_{j} q_{j} A^{(j)}_{\mu} \Bigr) \psi
 - m(\Phi) \bar{\psi} \psi
 \biggr]
\end{align}
where $\psi$ is a Dirac spinor of a test particle for the SWGC and has a charge $q_{i}$ under the gauge group 
$U(1)_{i}$ and a mass $m(\Phi)$,
and $\Phi^a$ is a real scalar field which may be different from $\phi^a$ in general.
Here, the covariant derivative $\nabla_\mu$ includes the spin connection.
The $\Phi^a$ is decomposed as
\begin{align}
 \Phi^{a} = \bar{\varphi}^{a} + \varphi^{a},
\end{align}
where $\bar{\varphi}^{a}$ is the background configuration of $\Phi^{a}$ and $\varphi^{a}$ denotes a fluctuation around 
the background.
With these, the mass $m(\Phi)$ is rewritten as
\begin{align}
 m(\Phi) = m ( \bar{\varphi} ) + \frac{ \partial m}{\partial \varphi^a} \varphi^a + \cdots.
\end{align}
$m(\bar{\varphi})$ is the mass of the $\psi$ in the background $\bar{\varphi}^{a}$, 
and the higher order terms of $\varphi^{a}$ give the interaction terms between $\varphi^{a}$'s and $\psi$.
Thus the Yukawa coupling reads:
\begin{align}
 S_{\mathrm{matter}} \supset \int d^4 x\, \sqrt{-g} \, \sum_{a} y_{a}(\bar{\varphi}) \varphi^{a} \bar{\psi} \psi, \\
 y_{a} (\bar{\varphi}) := \frac{ \partial m}{\partial \varphi^{a}} (\bar{\varphi}) = \partial_a m(\bar{\varphi}).
\end{align}
Then the SWGC for $\psi$ is given by
\begin{align}
 \sum_{i,j} f^{ij} q_{i} q_{j} \geq \frac{m^2}{2 M_{\mathrm{Pl}}^{2}} + \sum_{a,b} K^{ab} y_{a} y_{b},
\end{align}
where $f^{ij}$ and $K^{ab}$ are the inverse matrix of the $f_{ij}$ and $K_{ab}$ respectively.
This inequality can be interpreted as the total gauge repulsive force is stronger than the sum of
the attractive forces of the gravity and the total Yukawa interactions when 
we focus on forces acting between the test particle $\psi$:
$ |\vec{F}_{\mathrm{Coulomb}}| \geq   |\vec{F}_{\mathrm{gravity}}| +  |\vec{F}_{\mathrm{Yukawa}}|$.
The absolute value of long-range force mediated by massless fields in four dimension is expressed as
\begin{align}
 |\vec{F}| = \frac{A}{4\pi r^{2}},
\end{align}
where a numerator $A$ is the factor corresponding to each force: 
\begin{align}
 A_{\mathrm{Coulomb}} = \sum_{i,j} f^{ij} q_{i} q_{j},
 \quad
 A_{\mathrm{gravity}} = \frac{m^2}{2M_{\mathrm{Pl}}^{2}},
 \quad
 A_{\mathrm{Yukawa}} =  \sum_{a,b} K^{ab} y_{a} y_{b}
\end{align}
If the scalars $\varphi^a$ are heavy, Yukawa interactions are short-range forces
and neglected. Then the SWGC gets back to the WGC.

\subsection{Anomalous $U(1)$ symmetries}
\label{GSmech}

In this subsection, we review cancellation of chiral $U(1)$ gauge anomalies by axion fields.
In 4D effective field theories, gauge transformation of the axions can cancel the chiral anomalies produced by
light chiral fermions in the presence of topological terms of the gauge fields
and the St\"uckelberg couplings.
In field theories with an anomaly-free $U(1)$ gauge symmetry,
such axions are would-be Nambu-Goldstone bosons associated with the spontaneous breaking of the
$U(1)$ symmetry. After integrating out heavy fermions with chiral $U(1)$ charges, 
we can obtain anomalous $U(1)$ in the low energy limit \cite{Anastasopoulos:2006cz,Craig:2019zkf}.
In 4D string models, anomalies can be canceled by the Green-Schwarz mechanism involving string theoretic axions
that originate from tensor fields, when tadpoles of brane charges are canceled \cite{Dine:1987xk,Jockers:2004yj}.
We shall consider the 4D action involving axions
in addition to chiral fermions leading to chiral anomalies:
\begin{align}
 S_{\mathrm{axion}} =
 \sum_{i \in U(1)_{\mathrm{anomaly}}}  
 \int d^4 x 
 \biggl[
  - \frac{1}{2} \frac{m_i^2}{g_i^2 }\biggl( \sum_{I \in \mathrm{axions}} B_{iI} \partial_\mu \theta_I + A_\mu^{(i)}\biggr)^2
 + \sum_{I \in \mathrm{axions}}
 \frac{C_{iI} \theta_I}{32\pi^2 } \epsilon^{\mu \nu \rho \sigma} F_{\mu \nu}^{(i)} F_{\rho \sigma}^{(i)}
 \biggr].
\end{align}
Here, $\theta_I$ is an axion, $B_{iI}$ and $C_{iI}$ are constants, $m_i$ is the gauge boson mass. 
For the anomalous $U(1)$ symmetries, the fields transform as
\begin{align}
 \theta_I \to \theta_I - D_{Ii} \Lambda_i,
 \qquad
 A_\mu^{(i)} \to A_\mu^{(i)} + \partial_\mu \Lambda_i,
\end{align}
where $\Lambda_i$ is the transformation parameter, and we assume that 
$D_{Ii}$ satisfies
$\sum_I B_{iI} D_{Ij} = \delta_{ij}$.
The theory is invariant in the presence of chiral anomalies produced by gauge transformations 
against chiral fermions: 
\begin{align}
 S_{\mathrm{anomaly}} = 
 \sum_{i\in U(1)_{\mathrm{anomaly}}} \int d^4 x
 \bigg[
 \sum_{I \in \mathrm{axions}} \Lambda_i
 \frac{ C_{iI} D_{Ii}}{32\pi^2 } \epsilon^{\mu \nu \rho \sigma} F_{\mu \nu}^{(i)} F_{\rho \sigma}^{(i)}
 \bigg],
\end{align}
such that
$\delta_\Lambda S_{\rm total} = S_{\rm anomaly} + \delta_{\Lambda}S_{\rm axion} = 0$.
Thus, in terms of axions
the anomaly-free $U(1)$'s are determined such that the coefficients of $C_{iI}$'s are vanishing\footnote{
Once anomalous gauge fields are written as $A_{\mu}^{\mathrm{anomalous}} = \sum_i b_i A^{(i)}_\mu$,
$b_i$'s would be related to $c_i$'s in Eq.~(\ref{eq:keyeq}) through the orthogonality among $U(1)$'s.
If there exists a large hierarchy among $b_i$'s in $b_i (\partial^\mu \theta) A_{\mu}^{(i)}$,
some $c_i$'s would become very large.}.
4D effective action from 5D theory is also discussed, for instance, in Refs.~\cite{Lukas:1998tt, Cox:2019rro}.
The anomalous gauge bosons become massive as
\begin{align}
 -  \frac{1}{2} \frac{m_i^2}{g_i^2 } \bigl( B_{iI} \partial_\mu \theta_I + A_\mu^{(i)} \bigr)^2 =:
 -  \frac{1}{2} \frac{m_i^2}{g_i^2 } \bigl( \tilde{A}_\mu^{(i)} \bigr)^2,
\end{align}
after $\theta$'s are eaten by them as in spontaneous gauge symmetry breaking.
Further, for some non-anomalous gauge bosons, there can exist St\"uckelberg couplings
\begin{align}
 S_{\mathrm{axion}} =
 \sum_{i \in U(1)_{\text{non-anomalous}}}  
 \int d^4 x 
 \bigg[
 - \frac{1}{2} \frac{m_i^2}{g_i^2 } 
 \bigg( \sum_{I \in \mathrm{axions}} B'_{iI} \partial_\mu \theta_I + A_\mu^{'(i)} \bigg)^2 
 \bigg].
\end{align}
The non-anomalous gauge bosons can become massive as the anomalous ones.
Then, the repulsive forces mediated by such massive gauge bosons will not contribute to the WGC.
Hereafter, we suppose that this mechanism works in the quiver gauge theories studied in this paper,
and these terms are ignored otherwise stated.
%

\section{Quiver gauge theories and the WGC}
\label{Sec:quiver}
%
\begin{figure}[t]
\centering
\includegraphics[scale=0.34]{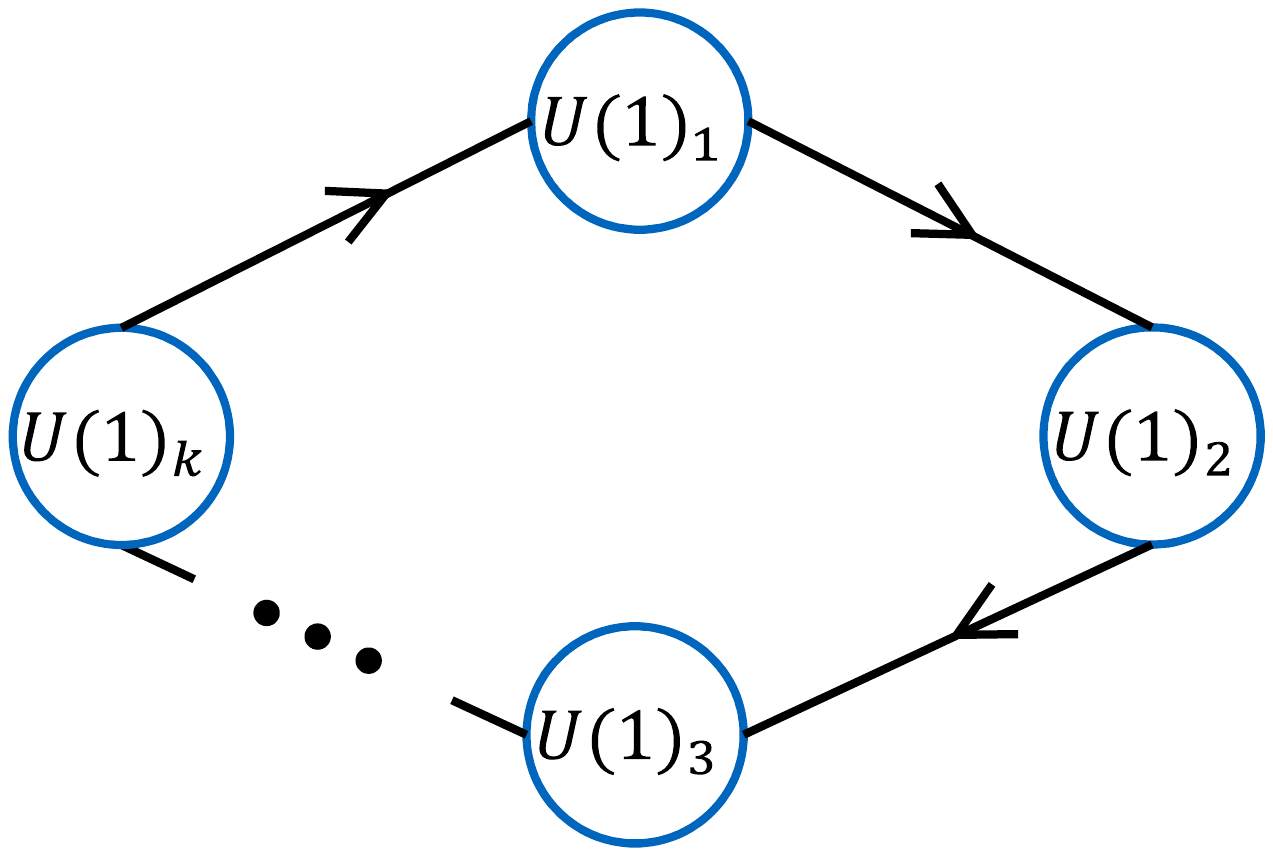}
\caption{
A quiver diagram with $k$ nodes.
}
\label{fig:quiverksimp}
\end{figure}
%

%
In this section, we discuss 
quiver theories with $U(1)^k$ gauge symmetry and identify anomaly-free gauge groups.
In general, computation of anomalies depends on the matter content in models.
To check anomaly-free $U(1)$'s systematically and identify the gauge couplings concretely, 
we focus on several types of models controlled by discrete symmetries.
However, an anomaly-free gauge coupling will be given by Eq.~(\ref{eq:keyeq}) in general cases.
As for a quiver diagram in this paper, 
each node implies a gauge group whereas each arrow among two nodes shows 
a left-handed chiral fermion charged under two gauge groups. 
The number of arrows shows that of matters and a direction of an arrow is corresponding to
the representation against two gauge groups.
An arrowhead corresponds to anti-fundamental representation while its opposite side means fundamental one.
For theories only with multiple $U(1)$ groups, 
(anti-)fundamental representation is supposed to have a charg	e $+1$ ($-1$).
A solid line shows a chiral (left-handed) fermion whereas a dashed line shows a complex scalar.
At first, we shall focus on non-supersymmetric gauge theories
with bi-fundamental chiral fermions of $(\bm{N}_{1} , \bar{\bm{N}_2})$ representation
under $ U(N_{1} )\times U(N_{2}) \times \cdots$ gauge group,
which is inspired by D-brane models.
Although there exist many types of quiver diagrams corresponding to gauge theories,
for simplicity we focus on theories including only $U(1)$ groups 
in the diagrams such as Fig.~\ref{fig:quiverksimp}. 
Since there exist chiral fermions, chiral gauge anomalies can generally be produced as a consequence.
We study cancellation condition of chiral anomalies to identify anomaly-free gauge couplings 
at the tree level, and apply the couplings to the WGC.
Anomaly-free conditions for $U(N)^{3}$ and $U(N)^{4}$ are discussed in Appendix~\ref{appA}.
For instance, in $SU(N)^k$ theories with a general $N$, 
non-abelian gauge anomaly cancellations require that the number of incoming arrows is equal to that of outgoing ones at each node.
In $U(1)^k$ theories we will simply mimic $SU(N)^k$ cases 
because in D-brane models a gauge group can be given by
$U(N) = U(1)\times SU(N)$ rather than just $SU(N)$, 
hence $U(1)$ and $SU(N)$ are considered simultaneously. 
We suppose that 
the anomalies are canceled as in Sec.~\ref{GSmech}
and then (non-)anomalous gauge fields get massive in a gauge invariant form.
Quiver gauge theories associated with deconstructed extra dimension \cite{ArkaniHamed:2001ca,Abe:2002rj}
could relate our approach to the weak coupling conjecture \cite{Buratti:2020kda}. 
In quiver gauge theories with $U(1)^k$ of our interest, the action is written by 
\begin{align}
 S = \sum_{j=1}^k \int d^4 x \biggl[
 - \frac{1}{4 g_j^2}  F^{(j)}_{\mu \nu} F^{(j) \mu \nu} +
\bar{\psi}_{j,j+1} i \gamma^{\mu} \bigl(\partial_\mu + i A^{(j)}_{\mu} - i  A^{(j+1)}_{\mu} \bigr) \psi_{j,j+1} 
+ \cdots \biggr],
\label{eq:Q-action}
\end{align}
where
ellipsis shows gravity and interaction terms among fermions which we have neglected.
We assume that kinetic mixings among gauge fields are absent at the tree level for simplicity,
and will ignore them in this paper.
The gauge field of $U(1)_j$ is denoted by $A^{(j)}_{\mu} $
and $\psi_{j,j+1}$ is a left-handed spinor with a charge of $(+1,-1)$ against
the $(U(1)_j, U(1)_{j+1})$ gauge group as noted above.
The index runs as $j=1,2,\ldots, k$ and satisfies $k+1 \equiv 1$. 
There will exist a symmetry\footnote{
See also Refs.~\cite{Gukov:1998kn,Burrington:2006uu,Garcia-Valdecasas:2019cqn}.
}
that shifts labels simultaneously as $j \to j +1$:
\begin{align}
g_j \to g_{j+1}, \quad A_{\mu}^{(j)} \to A_{\mu}^{(j+1)}, \quad \psi_{j,j+1} \to \psi_{j+1,j+2},
\end{align}
when we treat the gauge couplings as spurion fields, which are expected to be moduli fields in the string theory.
This can be regarded as a $\mathbb{Z}_{k}$ symmetry acting on $k$ nodes with a element of 
\begin{align}
 \left( \begin{array}{cccccc}
 0 & 1 & 0 & 0 & \cdots & 0 \\
 0 & 0 & 1 & 0 & \cdots & 0 \\
 0 & 0 & 0 & 1 & \cdots & 0 \\
  & & & & \ddots & \\
 1 & 0 & 0 & 0 & \cdots & 0
 \end{array}\right).
\end{align}
We can study anomalies and identify anomaly-free $U(1)$'s systematically owing to this symmetry as seen below.
This symmetry will be broken in the low energies when an 
anomaly-free gauge group is given by a linear combination of UV $U(1)$'s.
So, interactions of axions to gauge fields are expected to violate this discrete symmetry.
In terms of particle phenomenology, 
this theory may be the hidden sector for dark matter apart from the visible sector \cite{Costa:2020dph}.
In Appendix~\ref{Sec:appB}, we discussed also several quiver models not shown in this section.
%

\subsection{$U(1)^{2k-1}$}

We consider quiver gauge theories with $U(1)^{2k-1}$ groups as shown in Fig.~\ref{fig:quiverksimp}.
These types of (supersymmetric) models have often been studied in D-brane models on orbifolds or
intersecting/magnetized D-brane models.
They are used also to realize realistic Yukawa couplings or higher order couplings. 
We hereafter focus just on fermions producing anomalies.
As seen below, these theories can have an unique anomaly-free $U(1)$.

\subsubsection{$U(1)^3$}
\label{Sec3quiver}
%
\begin{figure}[t]
\centering
\includegraphics[scale=0.3]{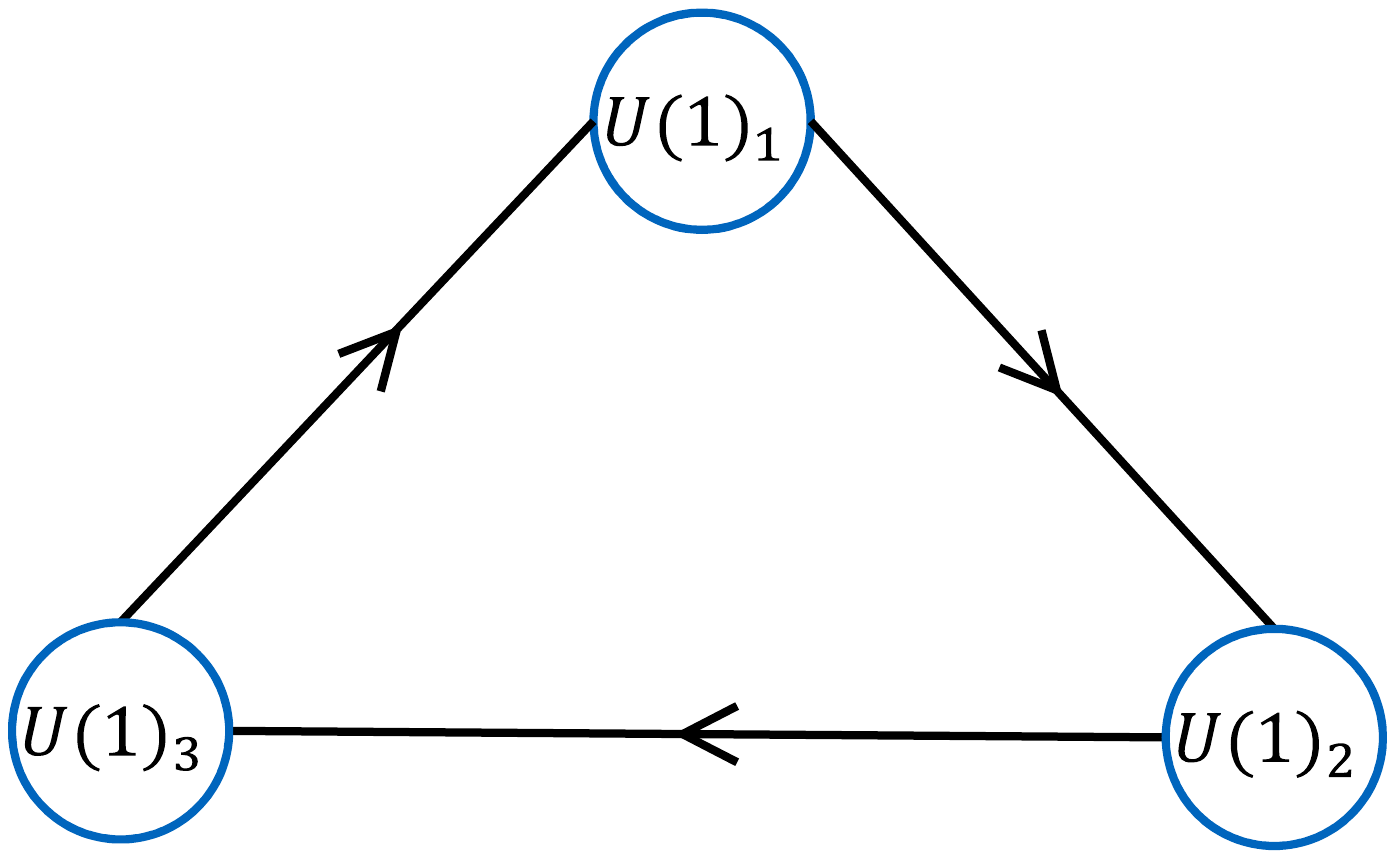}
\caption{
Three nodes quiver diagram.
}
\label{fig:quiverN3U1}
\end{figure}
%

%
One of the simplest case is the quiver gauge theory with $U(1)^3 = U(1)_1 \times U(1)_2 \times U(1)_3$ 
groups\footnote{
In a supersymmetric case, we have a Yukawa coupling. 
}
in Fig.~\ref{fig:quiverN3U1}.
As in Eq.~(\ref{eq:Q-action}), there exist three left-handed chiral fermions $\psi_{L i}~(i=1,2,3)$, which have 
charges of $(1,-1,0),~(0,1,-1)$ and $(-1,0,1)$ against $(U(1)_1,U(1)_2,U(1)_3)$ respectively\footnote{
These charge vectors will not satisfy the convex-hull condition if the same masses are given to the matters by hand. 
This is seen on the two dimensional section with two charge vectors whose magnitudes are $\sqrt{2}$
and between which the angle is $2\pi/3$. 
}. 
This model will have a ${\mathbb Z}_3$ symmetry as noted above,
and there is no other choices to connect each node.
The divergences of $U(1)^3$ chiral currents $j^{i\mu}~(i=1,2,3)$ are given by
\begin{align}
 \begin{cases}
 \partial \cdot j^{1} = Q_{2}- Q_{3}
 \\
 \partial \cdot j^{2} = Q_{3} -Q_{1}
 \\
 \partial \cdot j^{3} = Q_{1} - Q_{2},
 \end{cases}
\end{align}
where $\partial \cdot j^{i} = \partial_\mu j^{i \mu}$ and
$Q_{i}$ is the topological charge density, $Q_{i} = \frac{1}{32\pi^2} \epsilon^{\mu \nu \rho \sigma} 
F^{(i)}_{\mu \nu} F^{(i)}_{\rho \sigma}$.
Thus we define the anomaly-free $U(1)$ by
\begin{align}
 U(1)_{X} : = c_{1} U(1)_{1} + c_{2} U(1)_{2} + c_{3} U(1)_{3},
\end{align}
and impose the divergence of its current to vanish
\begin{align}
 \partial \cdot j^{X} = 
 \sum_{i = 1,2,3} c_{i} \partial \cdot j^{i} = 
 (-c_{2} +c_{3} ) Q_{1}
 + ( c_{1} - c_{3} ) Q_{2}
 + ( - c_{1} + c_{2} ) Q_{3}
 \equiv 0.
\end{align}
Then the solution is
\begin{align}
 U(1)_{X} = U(1)_{1} + U(1)_{2} + U(1)_{3}.
\end{align}
In this model, the anomaly-free gauge group is determined uniquely (up to overall normalization of the charges),
and its gauge coupling is given by
\begin{align}
\frac{1}{e_{X}^2} = \frac{1}{g_{1}^2} + \frac{1}{g_{2}^2} + \frac{1}{g_{3}^2}.
\end{align}
Here, the anomaly-free gauge coupling $e_X$ is written so that the gauge kinetic term becomes the canonical form:
\begin{align}
 - \sum_{j} \frac{1}{4g_{j}^2} F_{\mu\nu}^{(j)} F^{(j) \mu\nu} =
 -\frac{1}{4 e_{X}^2} F_{\mu\nu}^{(X)} F^{(X)\mu\nu} +
 ( \text{anomalous gauge fields}).
\end{align}
Thus, the anomaly-free gauge coupling $e_X$ can be smaller than the original $U(1)$ gauge couplings $g_i$'s.
It is noted that all the matters are then neutral under this anomaly-free $U(1)_{X}$, i.e.,  $\forall ~q_{X} =0$.
It seems that this model may not be naively applied to the WGC, but the presence of global symmetries is important.
The low energy Lagrangian will be given by
\begin{align}
 \mathcal{L} = \sum_{ i \in \text{ matter}} i \bar{\psi_{Li}} \Slash{\partial} \psi_{L i} 
 - \frac{1}{4 e_{X}^2} \bigl( F^{(X)}_{\mu\nu} \bigr)^2 + \cdots,
\end{align}
if anomaly-free gauge boson $A_{\mu}^{(X)}$ survives in low energy limit.
Ellipsis includes interactions among fermions and anomaly-free gauge boson
and there will additionally exist kinetic mixings such as $ K_{ij} \bar{\psi_{iL}} \Slash{\partial} \psi_{jL}$ and
Majorana mass terms of $ - M_{ij} \bar{\psi^C_{iL}} \psi_{jL}$ in low energy limit
after anomalous massive bosons are integrated out.
These terms will violate invariance under phase rotations of fermions.
Now the original ${\mathbb Z}_3$ symmetry acts as 
$e_X \to e_X,~A_\mu^{(X)} \to A_\mu^{(X)}$ and $\psi_{L i} \to \psi_{L i+1}$,
but whether this low energy theory has the ${\mathbb Z}_3$ symmetry depends on parameters for fermions.
Since all fermions are neutral under $U(1)_{X}$,
global symmetries will be hard to survive in the low energy limit
while discrete gauge symmetries originating from the anomalous $U(1)$'s can survive if any.
If global symmetries survive, this model is in the swampland.
It will be necessary to embed this model into string theory 
in order to know what kind of symmetries survives. 
This is beyond the scope of the paper and left for future work. 
%

\subsubsection{$U(1)^{2k-1}$}
%
\begin{figure}[t]
\centering
\includegraphics[scale=0.25]{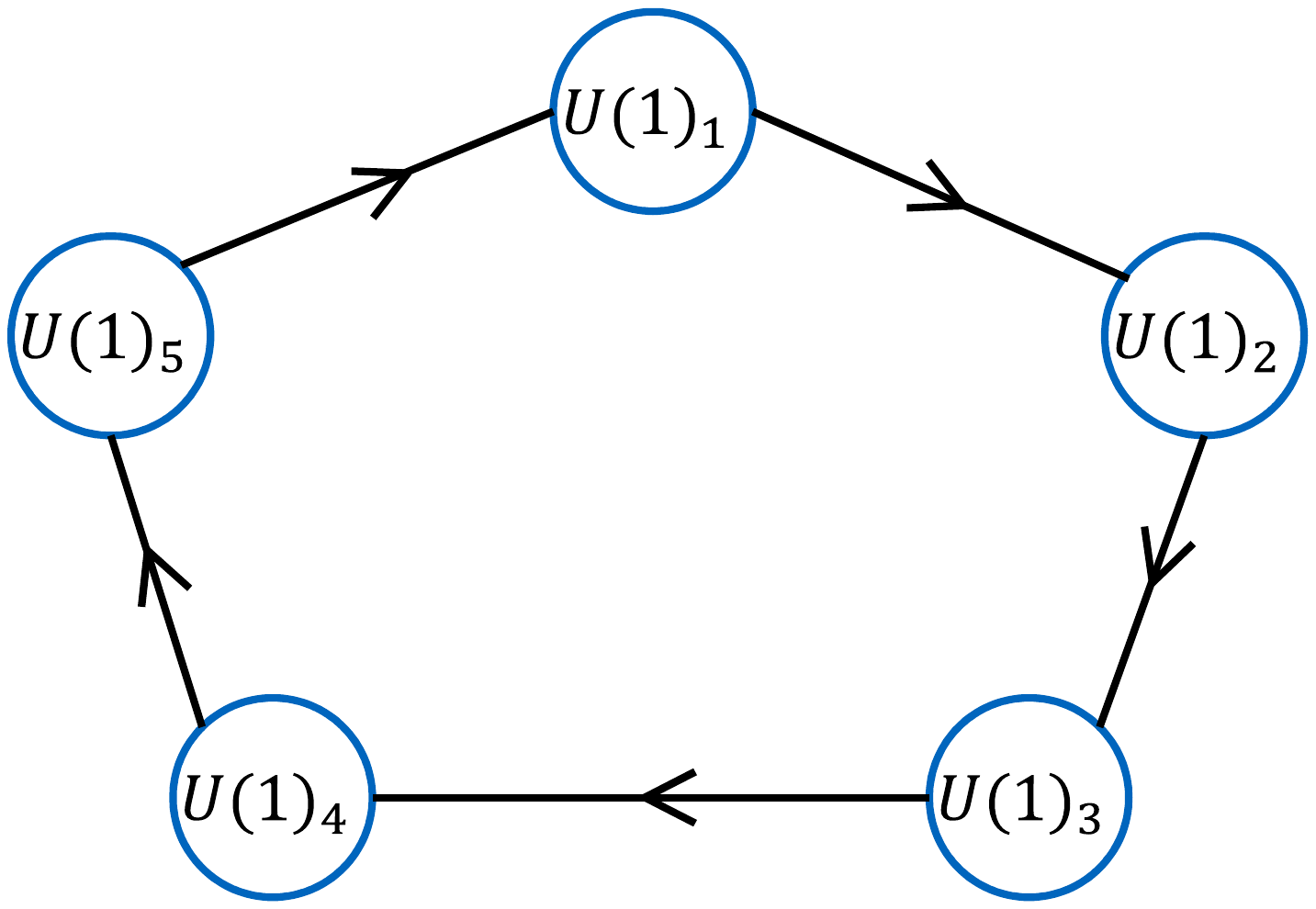}
\includegraphics[scale=0.25]{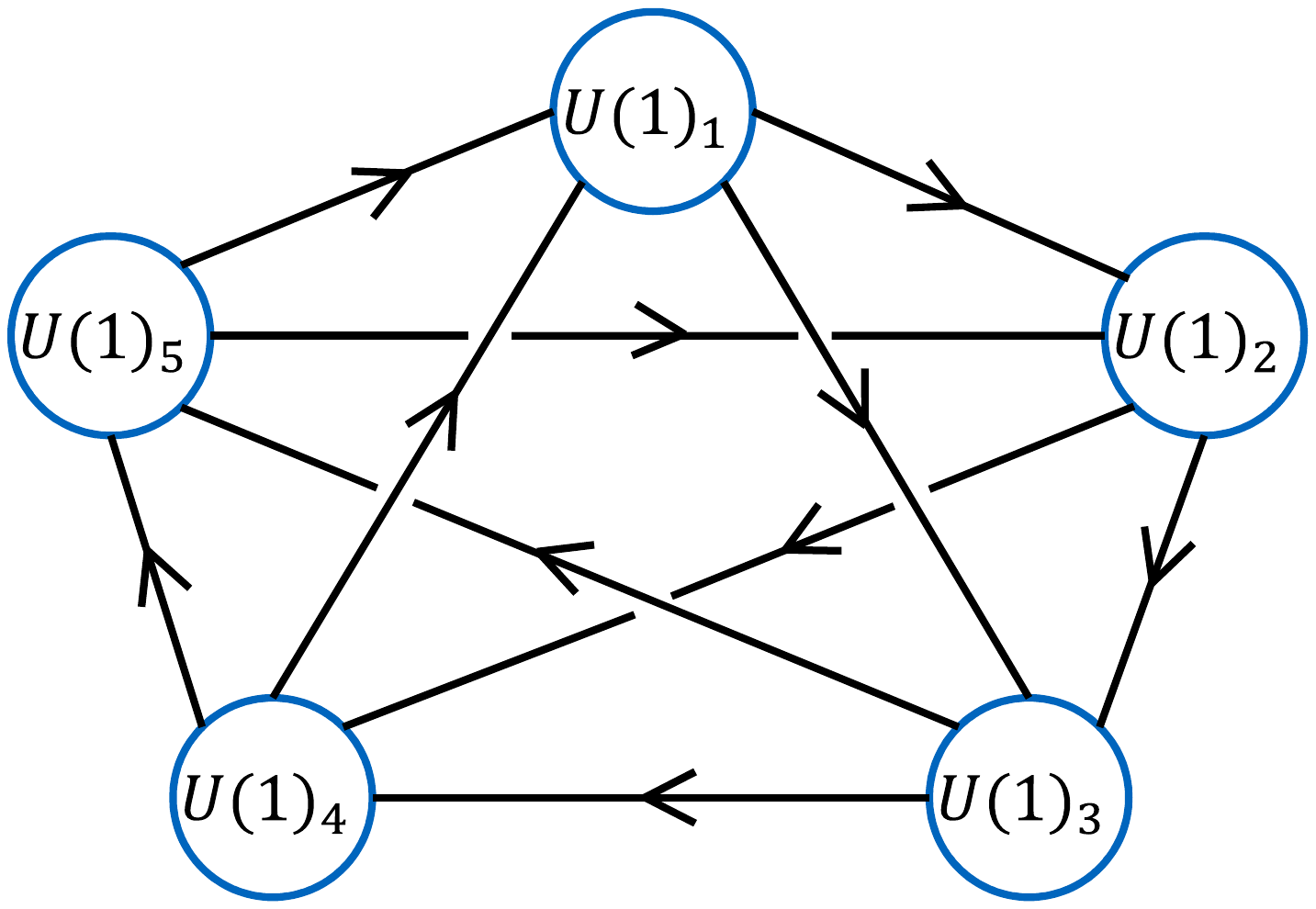}
\caption{
Quiver diagrams with five nodes.
Both diagrams have a $\mathbb{Z}_{5}$ cyclic symmetry among each node.
In the right diagram, all nodes are connected with arrows.
}
\label{fig:quiverN5simp}
\end{figure}

%
We consider quiver gauge theories with more general $U(1)^{2k-1}$ groups.
Fig.~\ref{fig:quiverN5simp} shows quiver diagrams with the five nodes,
and it is noted that the number of incoming arrows is equal to that of outgoing ones at each node
and both diagrams have a $\mathbb{Z}_{5}$ cyclic symmetry among each node.
As in Eq.~(\ref{eq:Q-action}) and in the left diagram of Fig.~\ref{fig:quiverN5simp}, 
we have five left-handed fermions charged 
against $(U(1)_1,U(1)_2,U(1)_3, U(1)_4, U(1)_5)$. 
The divergences of $U(1)^5$ chiral currents are given by
\begin{align}
 \partial \cdot
 \left( \begin{array}{c}
 j^{1} \\
 j^{2} \\
 j^{3} \\
 j^{4} \\
 j^{5}
 \end{array}\right)
 = 
 \left( \begin{array}{ccccc}
 0 & 1 & 0 & 0 & -1 \\
 -1 & 0 & 1 & 0 & 0 \\
 0 & -1 & 0 & 1 & 0 \\
 0 & 0 & -1 & 0 & 1\\
 1 & 0 & 0 & -1 & 0
 \end{array}\right)
 \left( \begin{array}{c}
 Q_{1} \\
 Q_{2} \\
 Q_{3} \\
 Q_{4} \\
 Q_{5}
 \end{array}\right).
\end{align}
The number of anomaly-free $U(1)$'s is given by that of zero eigenvalues of this coefficient matrix, 
and we find only one zero eigenvalue in this model.
The anomaly-free $U(1)$ is given by the corresponding eigenvector
\begin{align}
 U(1)_{\text{anomaly-free}} = U(1)_{1} + U(1)_{2} + U(1)_{3} +U(1)_{4} + U(1)_{5}.
 \label{eq:quiverN5simp}
\end{align}
Thus all matters are again neutral under this anomaly-free $U(1)$ and 
this system will not simply be applied to the WGC.
The situation is similar to the three quivers model in Sec.~\ref{Sec3quiver}.
The result is not changed by adding five chiral fermions to this model as 
in the right diagram of Fig.~\ref{fig:quiverN5simp}.
Then their action is additionally given by
\begin{align}
 S = \sum_{j=1}^5 \int d^4 x \biggl[
\bar{\psi}_{j,j+2} i \gamma^{\mu} \bigl(\partial_\mu + i A^{(j)}_{\mu} - i  A^{(j+2)}_{\mu} \bigr) \psi_{j,j+2} 
+ \cdots \biggr].
\end{align}
The anomaly coefficient matrix reads
\begin{align}
 \left( \begin{array}{ccccc}
 0 & 1 & 1 & -1 & -1 \\
 -1 & 0 & 1 & 1 & -1 \\
 -1 & -1 & 0 & 1 & 1 \\
 1 & -1 & -1 & 0 & 1\\
 1 & 1 & -1 & -1 & 0
 \end{array}\right).
\end{align}
Thus, the anomaly-free $U(1)$ is similary given by Eq.~(\ref{eq:quiverN5simp}).
So far we have discussed specific quiver models with three and five nodes, 
but the result can be simply extended to general models with odd number nodes as shown in Fig.~\ref{fig:quiverksimp}.
Since the entries of the anomaly coefficient matrix are composed of the same number of $1$ and $-1$ as above,
the anomaly-free $U(1)$ is uniquely determined as
\begin{align}
 U(1)_{\text{anomaly-free}} = \sum_{i}^{2k-1} U(1)_{i},
\end{align}
and the gauge coupling is given by
\begin{align}
\frac{1}{e^2} = \sum_{i=1}^{2 k -1 } \frac{1}{g_{i}^2}.
\end{align}
In concrete models, these are easily verified and 
it is checked also that the result does not change for models with odd nodes and full diagonal lines 
that are similar to the right diagram of Fig.~\ref{fig:quiverN5simp}.
As noted previously, however, there exist no charged chiral matters for this anomaly-free $U(1)$.

\subsubsection{$U(1)^{2k-1} \times U(1)^{2l-1} $ with vector-like matters}

%
\begin{figure}[t]
\centering
\includegraphics[scale=0.36]{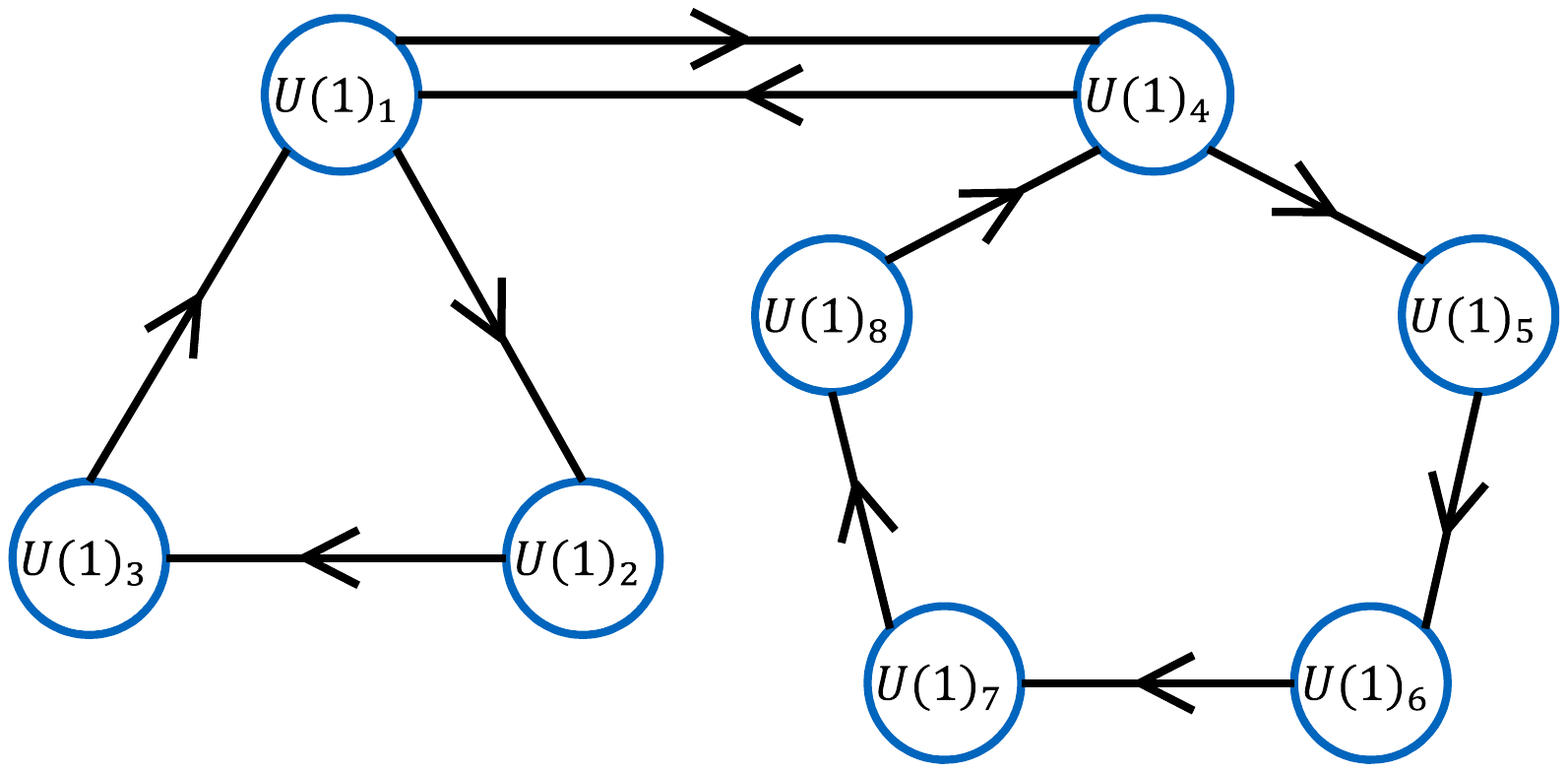}
\caption{
A quiver diagram of three nodes connected with five nodes by a pair of two arrows of vector-like matters.
}
\label{fig:quiver_three-five-2}
\end{figure}
%

%
We shall consider quiver gauge theories with $U(1)^{2k-1} \times U(1)^{2l-1} $ in the presence of vector-like matters.
As in Fig.~\ref{fig:quiver_three-five-2}, 
the correspoinding diagram is composed of two diagrams with odd nodes 
which are connected by a pair of two arrows of vector-like matters. 
Action is given by two kinds of Eq.~(\ref{eq:Q-action}) showing $U(1)^{2k-1} \times U(1)^{2l-1} $ symmetry
and vector-like part of
\begin{align}
 S = \int d^4 x \biggl[& 
 \bar{\psi}_{1,2k} i \gamma^{\mu} \bigl(\partial_\mu + i A^{(1)}_{\mu} - i  A^{(2k)}_{\mu} \bigr) \psi_{1,2k} 
 + \bar{\psi}_{2k,1} i \gamma^{\mu} \bigl(\partial_\mu - i A^{(1)}_{\mu} + i  A^{(2k)}_{\mu} \bigr) \psi_{2k,1} 
 \nonumber\\
 &- m \overline{\psi_{2k,1}^C}\psi_{1,2k}  + {\rm h.c.} \biggr],
\end{align}
where we assume that the bi-fundamental vector-like matters are charged under the gauge groups 
of $U(1)_1 \times U(1)_{2k}$ and that the mass $m$ remains non-zero in the weak gauge coupling limit.
In this case, the discrete symmetry is explicitly broken since  
$\psi_{1,2k}$ is transformed to $\psi_{2,2k+1}$ that is originally absent.
For instance, we focus on $U(1)^{3} \times U(1)^{5}$ theory with vector-like matter,
which is the case of $k=2$ and $l=3$.
Since vector-like matter does not contribute 
to the chiral anomalies, we have two anomaly-free $U(1)$'s as mentioned above: 
one denotes $U(1)_X$ from $U(1)^{3}$ and another denotes $U(1)_{X'}$ from $U(1)^{5}$.
Here,
\begin{align}
 U(1)_X = \sum_{i=1}^3U(1)_i, \qquad
 U(1)_{X'} = \sum_{i=4}^8 U(1)_i,
\end{align}
hence charges of vector-like matters are $(+1,-1)$ and $(-1, +1)$ for $(U(1)_X, U(1)_{X'})$
and other chiral matters are neutral for them.
The respective gauge couplings are given by 
\begin{align}
 \frac{1}{e_{X}^2} = \sum_{ i =1}^{3} \frac{1}{g_{i}^2},
 \qquad
 \frac{1}{e_{X'}^2} = \sum_{i =4}^{8} \frac{1}{g_{i}^2}.
\end{align}
These can be weaker than the original gauge couplings of $g_i$'s.
Then the WGC for the vector-like matter reads\footnote{
In the $(U(1)_X, U(1)_{X'})$ theory, we need also additional $(+1,+1)$ and $(-1, -1)$ vector-like matters
to satisfy the convex-hull condition so that extremal black holes with $(Q,Q)$ charge can decay.
This would imply that orientifold planes are required to cancel D-brane charges in the string theory.
However, our conclusion does not change.
}
\begin{align}
 e_{X}^2 + e_{X'}^2 \geq \frac{m^2}{2 M_{\mathrm{Pl}}^2}.
\end{align}
In the large limit of $k$ and $l$ with a given $m$ and $g_i$'s, we find 
\begin{align}
 e_{X}^2 + e_{X'}^2 \sim \frac{g_{i}^2}{k} + \frac{g_{i}^2}{l} \to 0,
\end{align}
and then the WGC can be violated since the couplings becomes very weak 
as long as the mass $m$ remains non-zero in the limit of $e_X \to 0$ and $e_{X'} \to 0$. 
Note that we now fix $g_i$'s but change only $k$ and $l$.
This indicates that there exists an upper bounds on the numbers of $U(1)$ gauge groups, $k$ and $l$ 
as $k +l \lesssim (g_i \frac{M_{\mathrm{Pl}}}{m})^2$
if the WGC is correct.

\subsection{$U(1)^{2k}$}
%
\begin{figure}[t]
\centering
\includegraphics[scale=0.3]{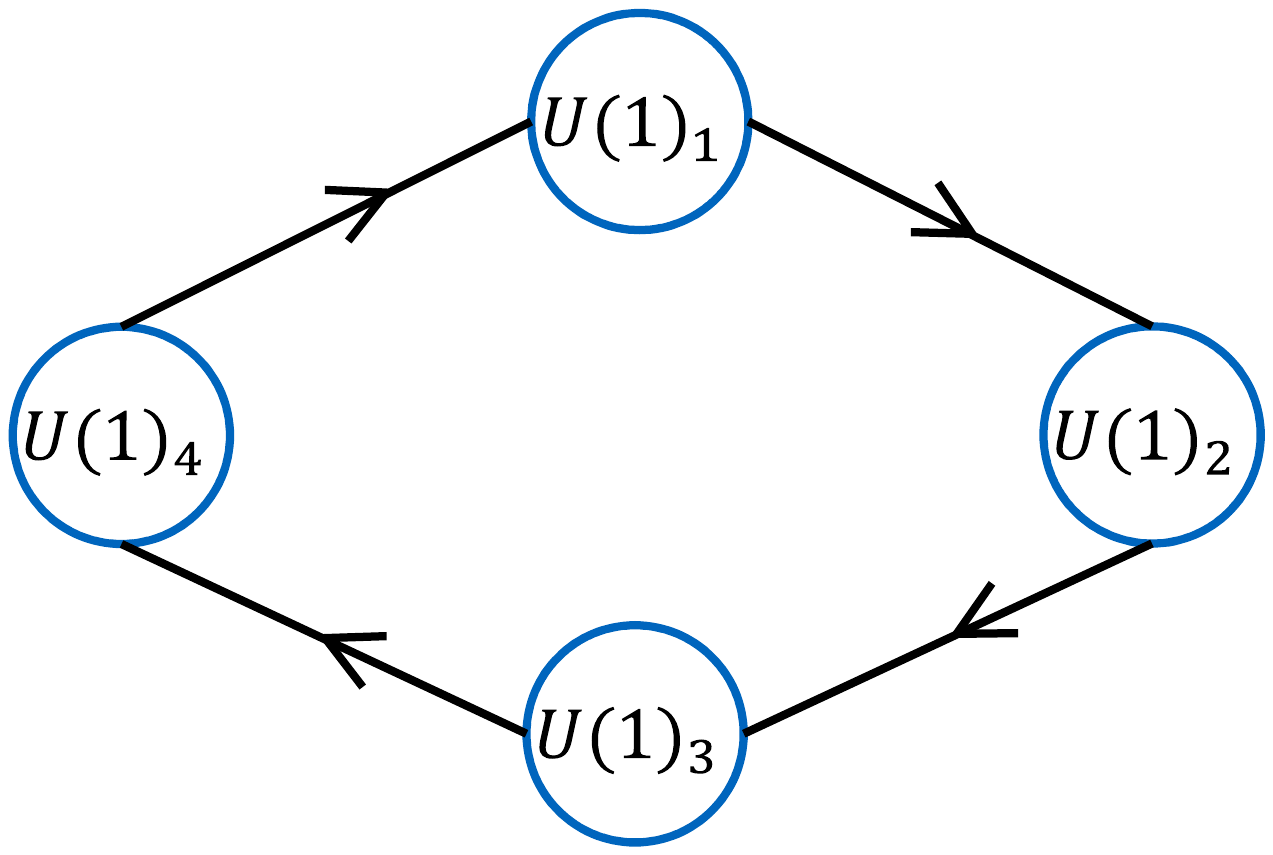}
\caption{
The $\mathbb{Z}_{4}$ symmetric quiver diagram with four nodes. 
}
\label{fig:quiverN4simp1}
\end{figure}
%

%
We consider quiver gauge theories with $U(1)^{2k}$ symmetry as shown in Fig.~\ref{fig:quiverksimp}.
These types of models are also studied in D-brane models similarly to $U(1)^{2k-1}$ cases.
As the simplest model with chiral anomalies, we focus on $U(1)^4$ symmetry and 
this model has four left-handed fermions as in Fig.~\ref{fig:quiverN4simp1}.
Divergences of each chiral current are given by
\begin{align}
 \partial \cdot
 \left( \begin{array}{c}
 j^{1} \\
 j^{2} \\
 j^{3} \\
 j^{4}
 \end{array}\right)
 = 
 \left( \begin{array}{cccc}
 0 & 1 & 0 & -1 \\
 -1 & 0 & 1 & 0 \\
 0 & -1 & 0 & 1 \\
 1 & 0 & -1 & 0
 \end{array}\right)
 \left( \begin{array}{c}
 Q_{1} \\
 Q_{2} \\
 Q_{3} \\
 Q_{4}
 \end{array}\right),
\end{align}
Since the anomaly coefficient matrix have two zero eigenstate,
this model has two independent anomaly-free $U(1)$'s, which are represented by the eigenvectors 
$(1,0 ,1 ,0 )^{\mathrm{T}}$ and $(0, 1, 0 , 1)^{\mathrm{T}}$.
The former relates first node to third one, whereas the latter does second node to fourth one.
The independent anomaly-free $U(1)$'s are generally given by
\begin{align}
 U(1)_{X} = & cU(1)_{1} + U(1)_{2} +c U(1)_{3} + U(1)_{4},
 \label{eq:U14U1X}
 \\
 U(1)_{X'} = & -\frac{1}{c} U(1)_{1} + U(1)_{2} - \frac{1}{c} U(1)_{3} + U(1)_{4},
 \label{eq:U14U1X'}
\end{align}
where $c$ is a free parameter that depends on the D-brane configuration in concrete UV string models \cite{Aldazabal:2000sa,Aldazabal:2000cn,Aldazabal:2000dg}\footnote{
A gauge boson in one of the two $U(1)$'s could be massive in UV models.
But, the behavior of gauge coupling will not change in the large $k$ limit. 
}, and will be a rational number. Otherwise, there exists a global symmetry \cite{Banks:1988yz,Banks:2010zn,Harlow:2018tng}.
The result does not change even if we add bi-fundamental vector-like matters that are charged 
under only $U(1)_1 \times U(1)_3$ or only $U(1)_2 \times U(1)_4$.
For $c = 0$, we find $U(1)_X = U(1)_2 + U(1)_4$ and $U(1)_{X'} = U(1)_1 + U(1)_3$.
It is noted that for a general $c$
a linear combination can violate the ${\mathbb Z}_4$ to ${\mathbb Z}_2^2$
exchanging $1 \leftrightarrow 3$ and $2 \leftrightarrow 4$.
The gauge couplings relevant to the anomaly-free $U(1)$'s read
\begin{align}
 \frac{1}{e_{X}^2} = & \frac{c^2}{g_{1}^2} + \frac{1}{g_{2}^2} + \frac{c^2}{g_{3}^2} + \frac{1}{g_{4}^2},
 \label{eq:U14gX}
 \\
 \frac{1}{e_{X'}^2} = & \frac{1/c^2}{g_{1}^2} + \frac{1}{g_{2}^2} +\frac{1/c^2}{g_{3}^2} + \frac{1}{g_{4}^2}.
 \label{eq:U14gX'}
\end{align}
In this model, the chiral fermions have non-trivial charges under these anomaly-free $U(1)$'s 
as shown in Table~\ref{tab:quiver_SWGC}. 
In the next section, we will numerically study the SWGC in this model by adding a complex scalar.

Extending this model to general theories with $U(1)^{2k}$ is simple,
and we can verify that there exists at least two anomaly-free $U(1)$'s in a concrete model.
So it is expected that in the large $k$ limit with a given $c$ and a fixed $g_i$,
anomaly-free gauge couplings become very small as in cases of $U(1)^{2k-1}\times U(1)^{2l-1}$.
Then there exists an upper bound on the number of abelian gauge groups if the WGC is correct
and a fermion mass remains non-zero in the large $k$ limit.
%

\section{A $U(1)^4$ model and the SWGC}
\label{Sec:4node}

In this section, we discuss the detail of $U(1)^4$ quiver gauge theory shown in the previous section
and its application to the SWGC at the tree level in the presence of a complex scalar field.
The motivation for this is to study SWGC in a more realistic (or string-inspired) model with a scalar field.
The SWGC shows numerically constraints of a smallness of gauge couplings against Yukawa couplings.
We also study a UV completion of 5D orbifold model for it.

\subsection{Constraints of the SWGC}
%
\begin{figure}[t]
\centering
\includegraphics[scale=0.3]{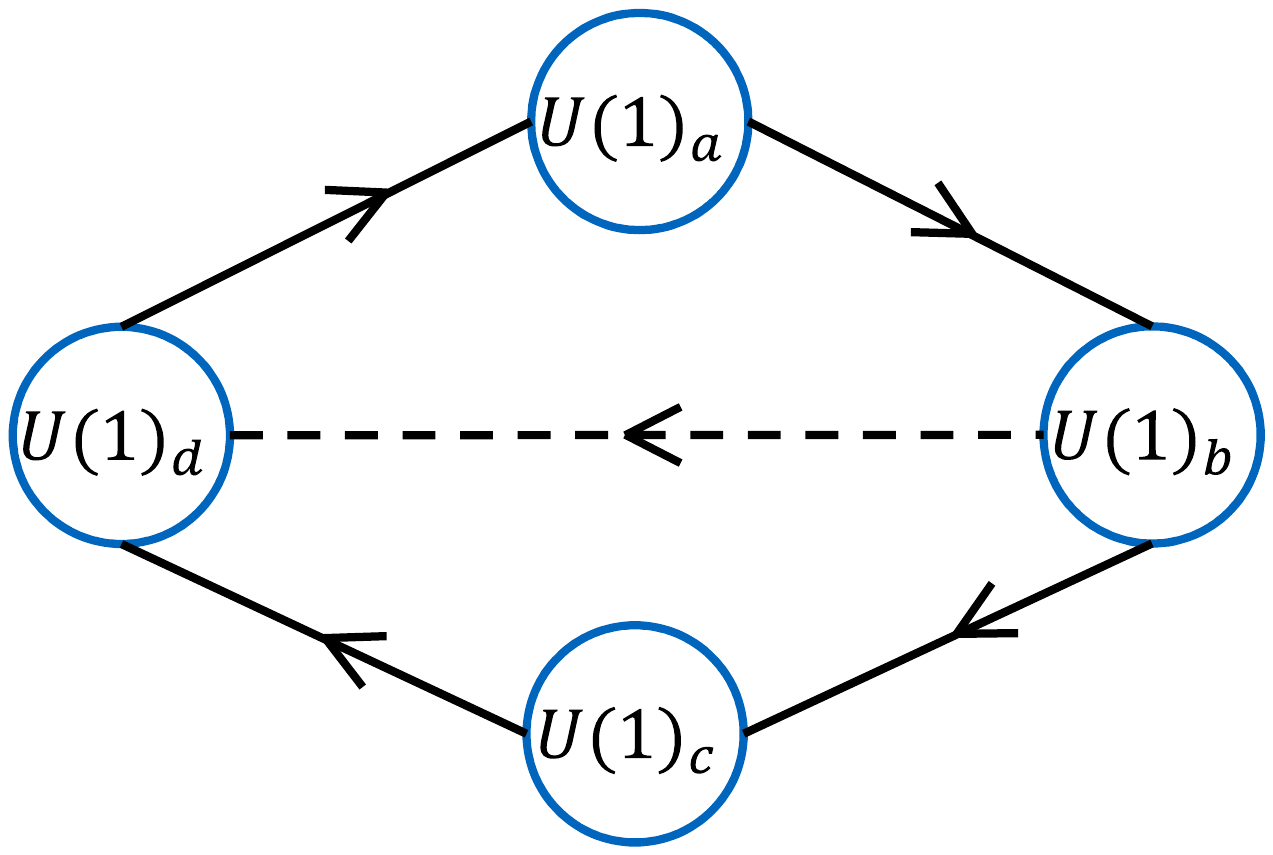}
\caption{
A quiver diagram of $U(1)^4$ model including a complex scalar. 
}
\label{fig:quiver_SWGC}
\end{figure}

%
Fig.~\ref{fig:quiver_SWGC} shows a quiver diagram of 
$U(1)^4$ model in the presence of a complex scalar $\varphi$, whose 
charge is $(+1,-1)$ for $(U(1)_b,U(1)_d)$\footnote{
The direction of the dashed arrow shows the scalar charge same as chiral fermions.
We changed the names of gauge groups from $U(1)_1 \times U(1)_2 \times U(1)_3 \times U(1)_4$ 
to $U(1)_a \times U(1)_b \times U(1)_c \times U(1)_d$ and accordingly those of left-handed fermions
from $(\psi_{1,2},~\psi_{2,3},~\psi_{3,4},~\psi_{4,1})$ to $(\psi_{ab},~\psi_{bc},~\psi_{cd},~\psi_{da})$
for the latter convenience.}.
Due to this scalar field, we have Yukawa couplings of
\begin{align}
 \mathcal{L}_{\rm Yukawa}=
 - y \varphi \overline{\psi_{ab}^C}\psi_{da}
 - y' \varphi^\dag \overline{\psi_{bc}^C}\psi_{cd} + \mathrm{h.c.}
\end{align}
This model is inspired by intersecting brane models \cite{Ibanez:2001nd,Cremades:2002cs}.
No $\mathbb{Z}_4$ symmetry exists.
This is because $\varphi$ can be written as $\varphi_{bd}$ in the view point of the $U(1)$ charges
and hence $\varphi_{bd}$ is transformed to $\varphi_{ca}$ that is originally absent.
There could exist $\mathbb{Z}_2$ that simultaneously exchanges the labels as $a \leftrightarrow c$ and $b \leftrightarrow d$ for $y = y'$, 
if we can identify $\varphi_{bd}=\varphi_{db}^\dag$.
As seen in the previous section, two anomaly-free $U(1)$'s are given by
\begin{align}
 U(1)_{X} =& c U(1)_{a} + U(1)_{b} + cU(1)_{c} +  U(1)_{d},
 \\
 U(1)_{X'} = & -\frac{1}{c} U(1)_{a} +U(1)_{b} -\frac{1}{c} U(1)_{c} + U(1)_{d},
\end{align}
where a free parameter $c$ is a rational number and can be fixed in concrete models by the brane configuration in the string theory.
The charges of the fields are summarized in Table~\ref{tab:quiver_SWGC}.
\begin{table}[t]
\centering
 \begin{tabular}{|c||c||c|}\hline
 Fields & $q_{X}$ & $q_{X'}$   \\ 
 \hhline{|=#=#=|}
 $\psi_{ab}$ & $-1 +c$ & $-1 -1/c$ \\ \hline
 $\psi_{da}$ & $1- c$ & $1+1/c$ \\ \hline
 $\psi_{bc}$ & $1-c$  & $1+ 1/c$  \\ \hline
 $\psi_{cd}$ & $-1 + c$ & $-1 -1/c$ \\ \hline
 $\varphi$ & $ 0 $ & $ 0 $ \\ \hline
 \end{tabular}
\caption{
The charges of fields for the anomaly-free $U(1)_{X} \times U(1)_{X'}$ group.
}
\label{tab:quiver_SWGC}
\end{table}
The effective Lagrangian showing two anomaly-free $U(1)$'s may read
\begin{align}
 \mathcal{L} = &\sum_{I= ab,bc,cd,da} i \bar{\psi_{I}} \Slash{D} \psi_{I} - |\partial_\mu \varphi|^2
 - \frac{1}{4 e_{X}^2} \bigl( F^{(X)}_{\mu\nu} \bigr)^2  - \frac{1}{4 e_{X'}^2} \bigl( F_{\mu\nu}^{(X')} \bigr)^2
 \nonumber\\
 & - \bigl[ y \varphi \bar{ \psi_{ab}^C} \psi_{da} 
 + y' \varphi^{\dagger}  \bar{\psi_{bc}^{C}} \psi_{cd} + \mathrm{h.c.}]  + \cdots,
 \label{eq:4-nodes}
\end{align}
where $D_\mu = \partial_\mu + i q_{X} A_\mu^{(X)} + i q_{X'} A_\mu^{(X')}$,
gauge bosons relevant to two anomalous $U(1)$'s are neglected since they become massive 
if the Green-Schwarz mechanism works.
Yukawa couplings between the complex scalar and chiral fermions are denoted by $y$ and $y'$, 
which are not the same in general.
It is noted that vector-like pairs of $\psi_{ab} + \psi_{da}$ and $\psi_{bc} + \psi_{cd}$ will 
constitute the Dirac spinors\footnote{
From the view point of the anomaly-free $U(1)$'s, there may exist
Yukawa couplings including $\bar{ \psi_{ab}^C} \psi_{bc} $ and $\bar{ \psi_{cd}^C} \psi_{da} $,
but they are supposed to be much smaller than $y$ and $y'$ here.
Similarly, there may exist Dirac masses to these fermions because $\varphi$ is singlet for the anomaly-free $U(1)$'s
in the low energy, but the masses would be negligibly small against the Planck scale.
}. 
Now scalar $\varphi$ is neutral under anomaly-free $U(1)$'s and will not be considered for the SWGC.
The scalar potential will be neglected hereafter with an assumption that $\varphi$ is sufficiently light
at energy scales of our interest 
since the scalar potential will be model-dependent. 
To check strong SWGC \cite{Gonzalo:2019gjp} for $\varphi$ is an interesting issue, but this is left for future work
and we focus on the SWGC for fermions with non-trivial anomaly-free gauge charges.
Here the anomaly-free gauge couplings are given by
\begin{align}
 \frac{1}{e_{X}^2} = & \frac{c^2}{g_{a}^2} + \frac{1}{g_{b}^2} + \frac{c^2}{g_{c}^2} + \frac{1}{g_{d}^2},
 \label{eq:couplingU(1)X}
 \\
 \frac{1}{e_{X'}^2} = & \frac{ 1/c^2}{g_{a}^2} + \frac{1}{g_{b}^2} +\frac{1/c^2}{g_{c}^2} + \frac{1}{g_{d}^2}.
 \label{eq:couplingU(1)X'}
\end{align}
In the presence of a very light $\varphi$, the SWGC can be expressed as
\begin{align}
 (-1 + c)^2 e_{X}^2 + \biggl( 1 + \frac{1}{c} \biggr)^2 e_{X'}^2 \geq \frac{M^2}{2 M_{\mathrm{Pl}}^2} + \frac{Y^2}{2},
 \label{eq:general_exSWGC}
\end{align}
for a test fermion. 
Here, $Y=y$ and $M=y{\rm Re}(\varphi)$ for a Dirac fermion of $\psi_{ab} + \psi_{da}$, 
whereas $Y=y'$ and $M=y'{\rm Re}(\varphi)$ for $\psi_{bc} + \psi_{cd}$. 
A factor $Y^2/2$ is obtained because of the canonical normalization of Re$(\varphi$),
and Im$(\varphi$) contributes to the spin-dependent interaction that is not $1/r^2$-force.
If the scalar is sufficiently heavy, $Y$ does not contribute to the SWGC conditon owing to exponentially damping force
and hence the WGC can be easily satisfied.
To reduce the number of parameters, we will set $g_b = g_d =: g$ for simplicity.
In the next subsection, we will study this situation realized in the 5D orbifold model.
Thus this equation can be rewritten as
\begin{align}
\frac{(1-c)^2}{c^2 ( g^2/ g_{a}^2 + g^2/g_{c}^2) +2 } 
+ \frac{ (1 + 1/c)^2}{(1/c^2) (g^2 /g_{a}^2 +g^2 /g_{c}^2) +2 } \gtrsim \frac{1}{2}\bigg(\frac{Y}{g}\bigg)^2.
 \label{eq:exSWGC}
\end{align}
Here, the masses are neglected because $M/M_{\rm Pl} \ll 1$ is numerically expected 
in the effective field theory.
Indeed, there is almost no change in appearance of the plots for $M/g M_{\rm Pl} \lesssim 0.1$,
where $g M_{\rm Pl}$ is expected as a cutoff scale \cite{ArkaniHamed:2006dz}, when the scalar is massless.
It is noted also that a gauge boson in either $U(1)_X$ or $U(1)_{X'}$ may be massive
owing to the St\"uckelberg coupling and then either $e_X$ or $e_{X'}$ vanishes in Eq.~(\ref{eq:exSWGC}).
%

%
\begin{figure}[t]
\centering
\includegraphics[scale=0.21]{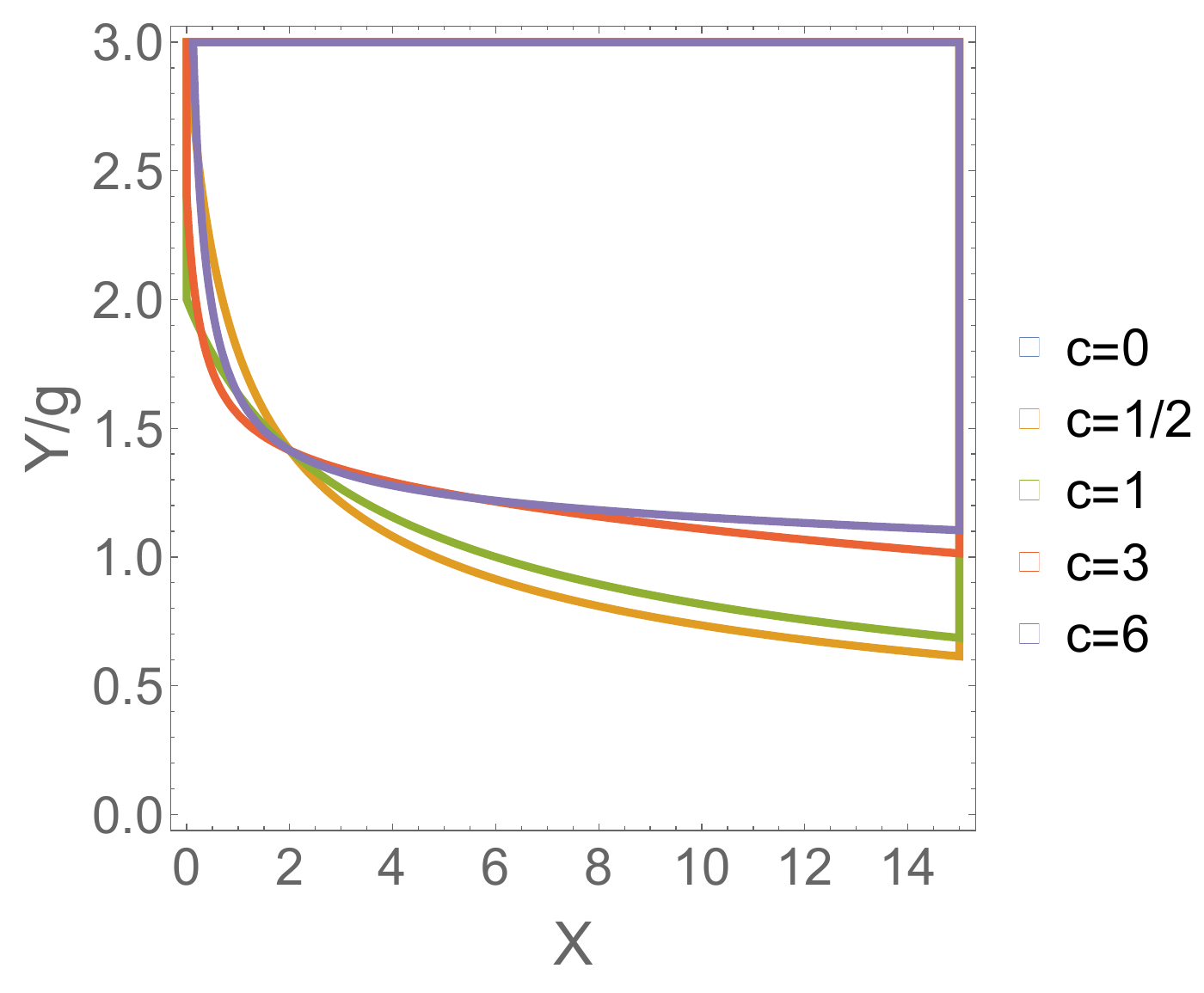}
\includegraphics[scale=0.21]{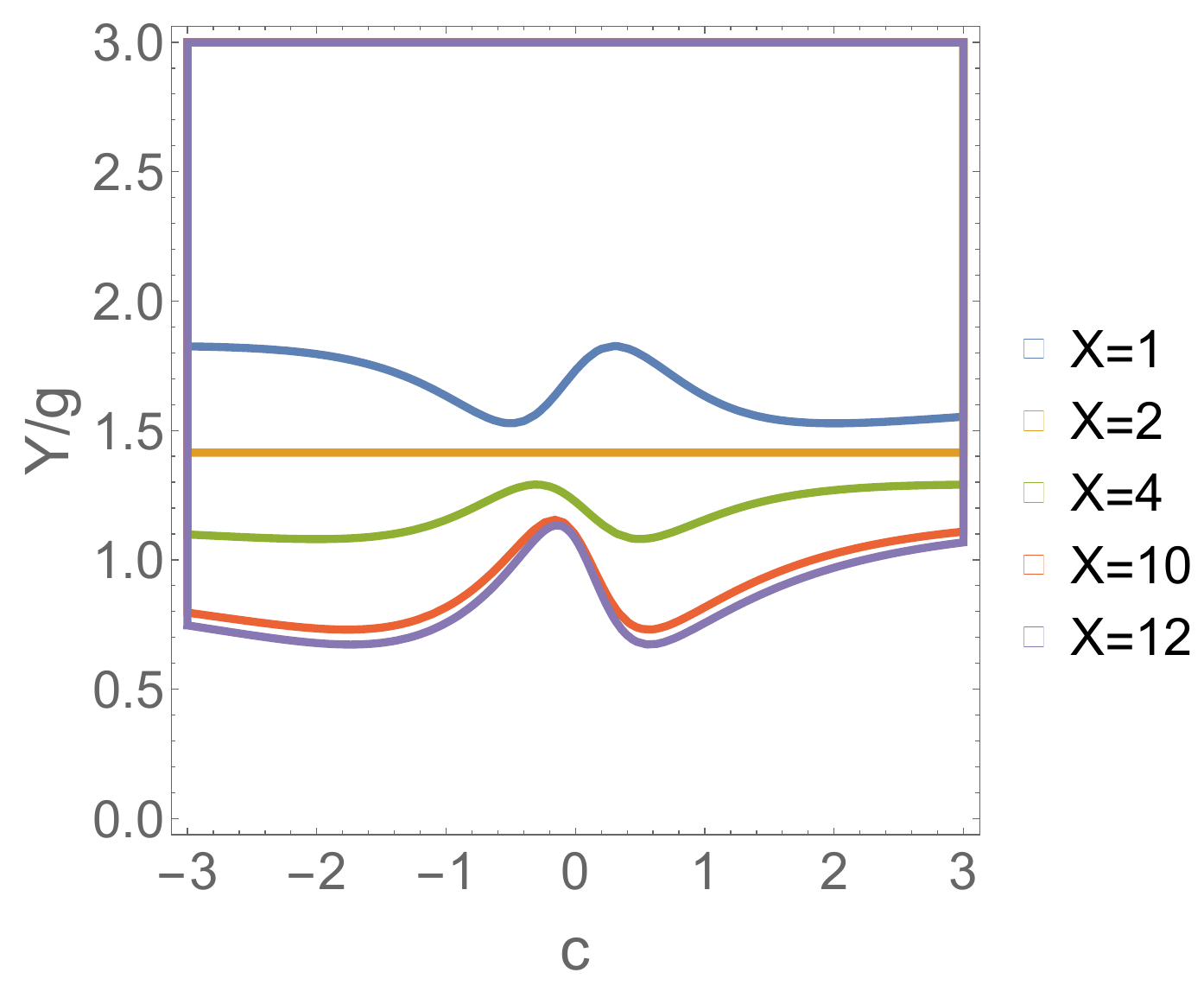}
\includegraphics[scale=0.21]{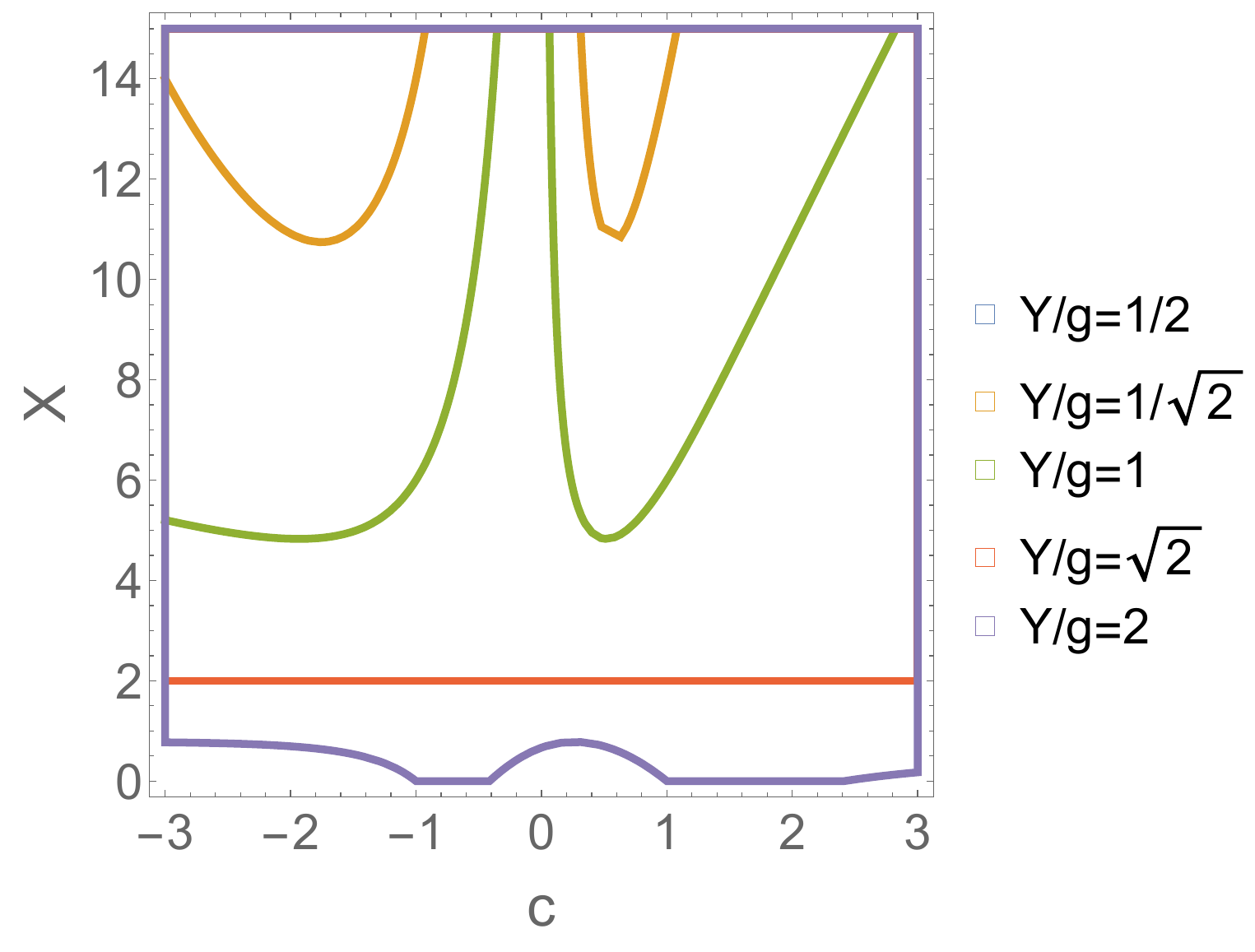}
\\
\includegraphics[scale=0.21]{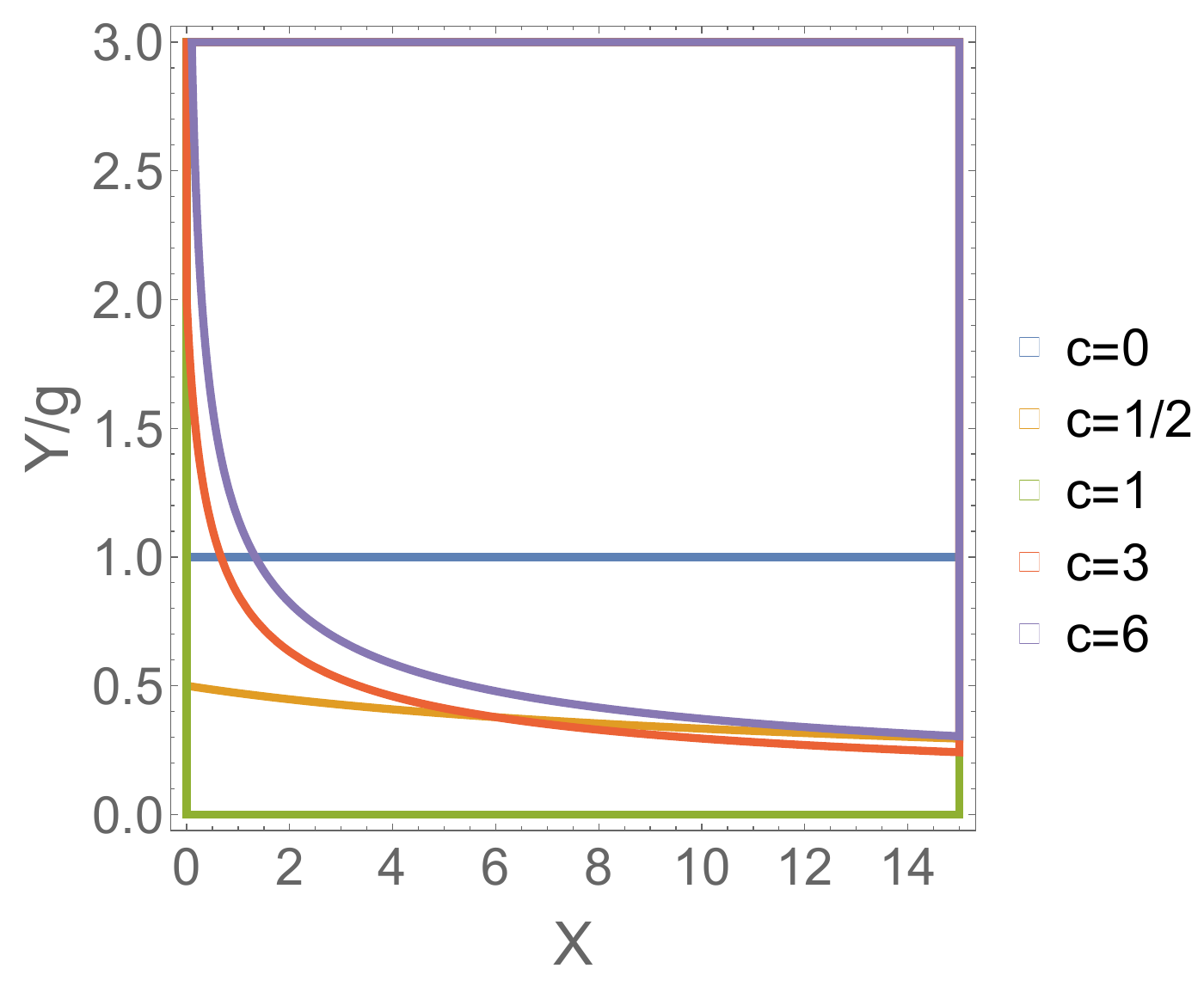}
\includegraphics[scale=0.21]{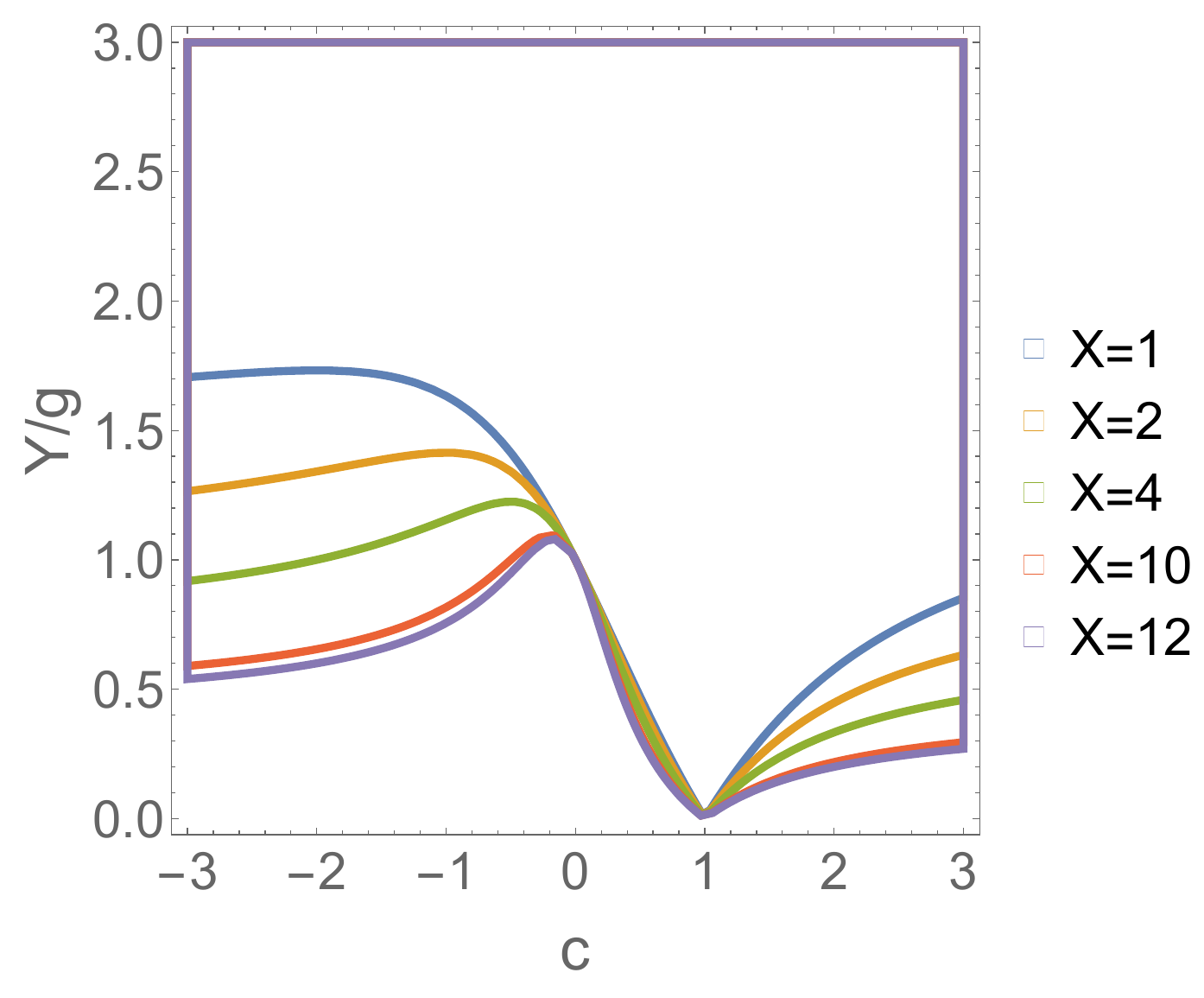}
\includegraphics[scale=0.21]{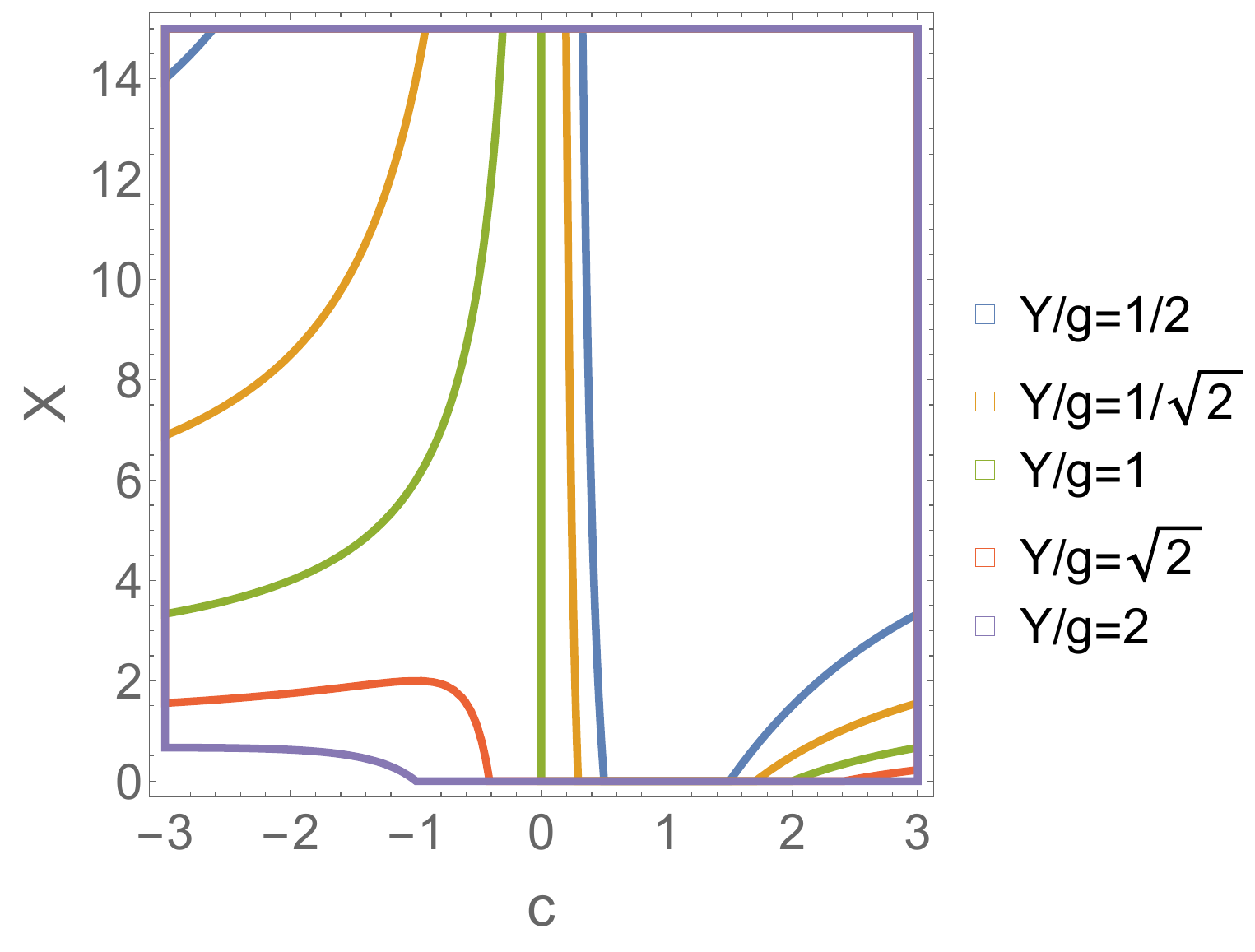}
\\
\includegraphics[scale=0.21]{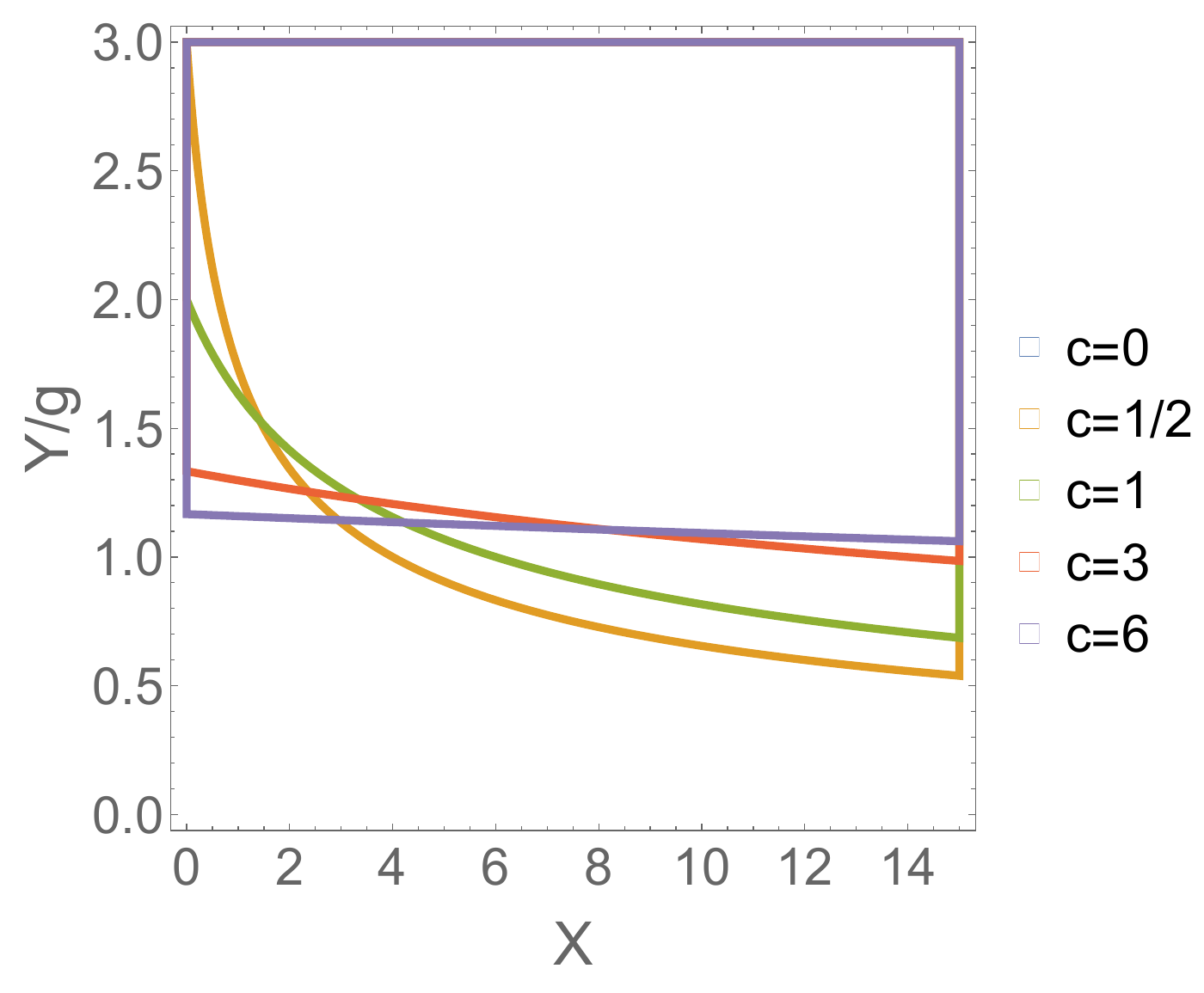}
\includegraphics[scale=0.21]{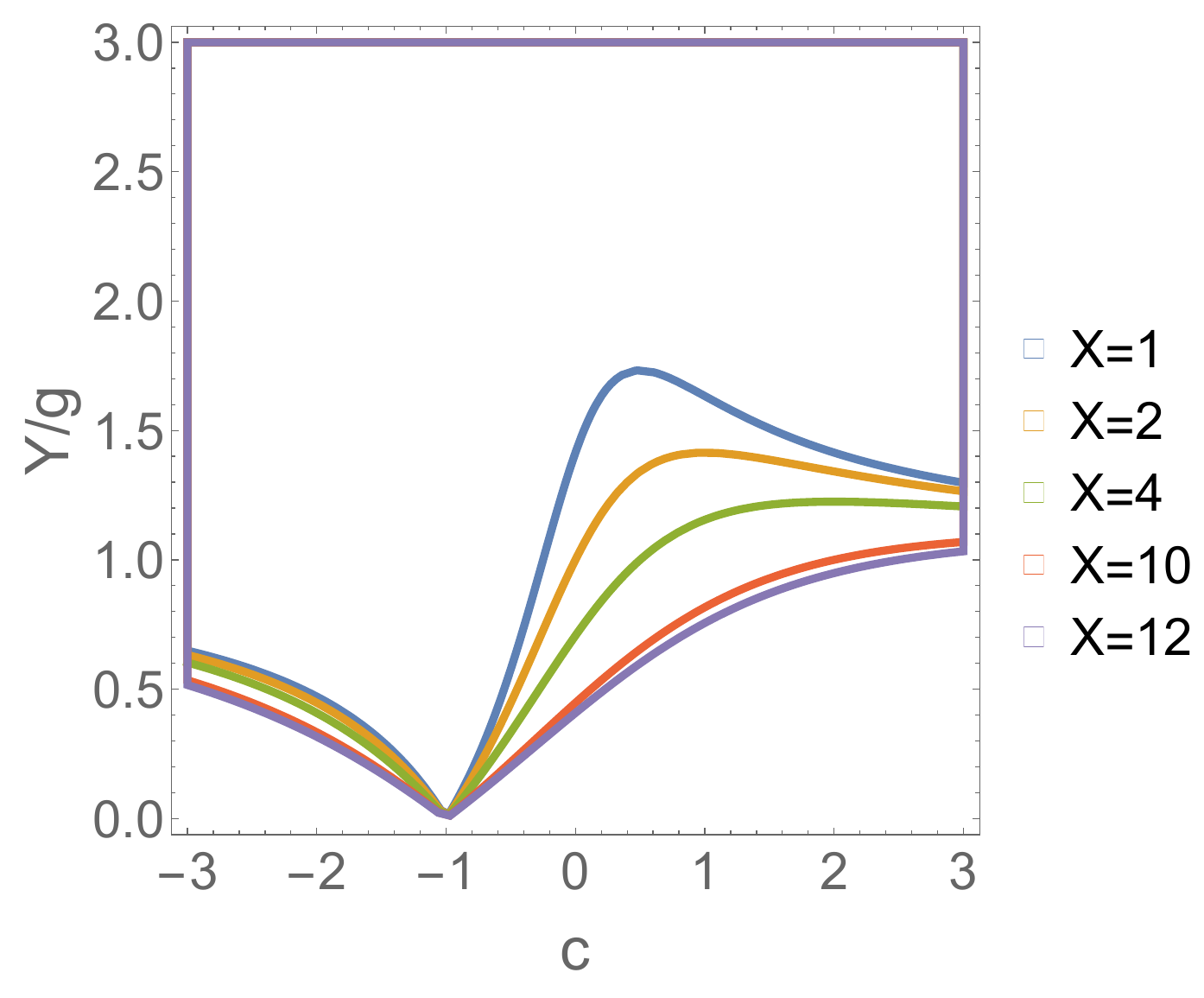}
\includegraphics[scale=0.21]{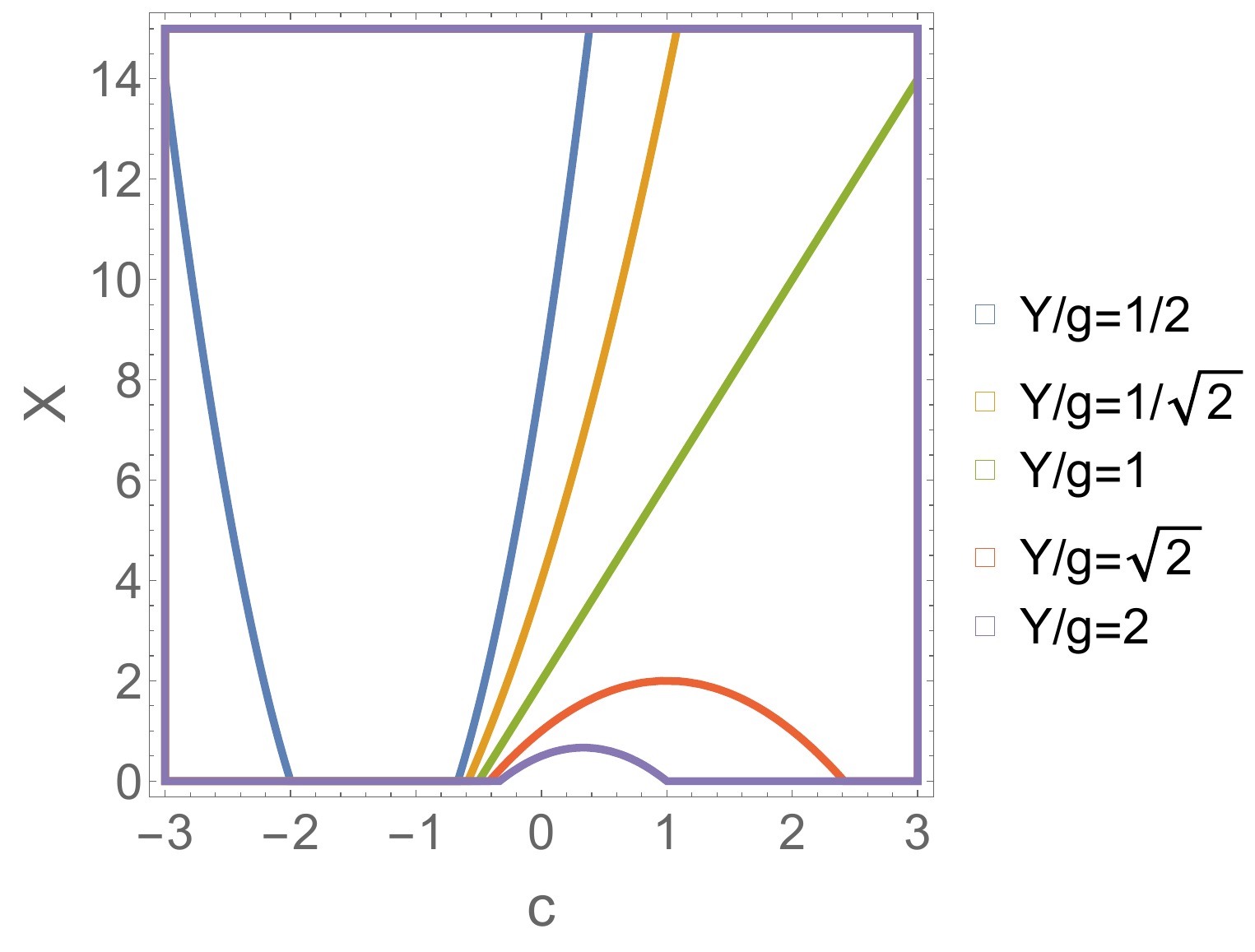}
\caption{
Plots of the SWGC constraints in $(X, Y/g)$-, $(c, Y/g)$- and $(c, X)$-planes.
The top panels plot the constraints of Eq.~(\ref{eq:exSWGC}).
The middle (bottom) figures show the similar plots with $e_{X'} =0$ 
($e_{X} =0$), when $U(1)_{X}$ ($U(1)_{X'}$) gauge group survives in low energy limit.
The condition is saturated on each lines, below which there exist an allowed region.
In the presence of mass, the SWGC is violated on each line.
}
\label{fig:testplot}
\end{figure}
%

%
In the top panels of Fig.~\ref{fig:testplot}, we show the plots of the SWGC (\ref{eq:exSWGC}) 
in the $(X, Y/g)$-, $(c, Y/g)$- and $(c, X)$-planes,
where $X := g^2/g_{a}^2 + g^2/g_{c}^2$. The each line saturates Eq.~(\ref{eq:exSWGC}),
hence the allowed region exists below them.
In the presence of the mass, the SWGC is violated on each line.
Note that these plots are symmetric under $c \to -1/c$ owing to the definition of the anomaly-free $U(1)$'s.
A region for a large Yukawa coupling is excluded by the SWGC.
For a large $X$, the constraint becomes tighter. In other words, a big discrepancy between gauge couplings is disfavored.
It turns out that the constraint becomes stronger near $c = \pm1$ because either $e_X$ or $e_{X'}$ vanishes then.
We find also that the constraint is independent of $c$ for special values of $X = 2$ and $Y/g = \sqrt{2}$.
This is because for $X=2$ the left hand side of Eq.~(\ref{eq:exSWGC}) becomes unity
and hence for $Y/g \geq \sqrt{2}$ the SWGC is then violated in the presence of the mass term.
We can find also that the constraint becomes weaker as the $c>0$ increases 
in the top-left panel for $X>2$, because either $e_{X}$ or $e_{X'}$ gets stronger then
whereas the constraint does not depend on $c$ for $X\ll 1$.
It is noted that in the string theory $c$ depends on the D-brane configuration and 
the couplings depend on moduli fields with the fixed configuration.
The middle (bottom) figures show the similar plots with $e_{X'} =0$ 
($e_{X} =0$), when only a gauge boson of $U(1)_{X}$ ($U(1)_{X'}$) remains massless 
and mediates the long-range repulsive force.
In these cases, the condition of the SWGC tends to give tighter constraints.
%

\subsection{A $U(1)^4$ model from $S^1 / \mathbb{Z}_2$ orbifold and the SWGC}
\label{sec:orbifoldmodel}
%
\begin{table}[t]
\centering
 \begin{tabular}{|c||c|c|c|} \hline
 Fields & $U(2)$ & $U(1)_{a}$ & $U(1)_{c}$ \\
 \hhline{|=#=|=|=|}
 $A_{M}$ & adj & $0$ & $0$ \\ \hline
 $\Psi_{a}$ & $\bm{2}_{1/2}$ &$ -1$ & $0$ \\ \hline
 $\Psi_{c}$ & $\bar{\bm{2}}_{-1/2}$ & $0$ & $+1$ \\ \hline
 $A^{(a)}_{M}$ & $0$ & adj & $0$ \\ \hline
 $A^{(c)}_{M}$ & $0$ & $0$ & adj \\ \hline
 \end{tabular}
\caption{
Table of the field contents and their charges in 5D model 
for realizing $U(1)^4$ gauge theory in 4D.
Subscripts of $U(2)$ representation for fermions are $U(1)$ charges against the overall $U(1) \in U(2)$. 
}
\label{tab:toymodel}
\end{table}
%

%
We consider a 5D gauge theory with $U(2) \times U(1)_a \times U(1)_c$ on the $S^1 / \mathbb{Z}_2$ orbifold
for realizing chiral fermions.
The purpose of this subsection is to give a concrete Yukawa coupling associated with the gauge coupling
and a relation between gauge couplings as in the previous subsection via 
the symmetry breaking of $U(2) \to U(1)_b \times U(1)_d$ by an orbifold projection.
The fields contents and their representations are exhibited in Table~\ref{tab:toymodel}.
The 5D action is given by
\begin{align}
 S_{\text{5D}} = \int_{M_4 \times S^1 / \mathbb{Z}_2} d^4x dy \, \sqrt{ -G_{5}} \biggl[ &
 \frac{1}{2 \kappa_{5}^2} \mathcal{R}_{5} - \frac{1}{2 \hat{g}_{2}^2} \tr ( F_{MN} )^2
 - \frac{1}{4 \hat{g}_{a}^2} \bigl( F^{(a)}_{MN} \bigr)^2 - \frac{1}{4 \hat{g}_{c}^2} \bigl( F^{(c)}_{MN} \bigr)^2 
 \nonumber\\
 & + \bar{\Psi_{a}} ( i \Slash{D} -M_{a} ) \Psi_{a}
 + \bar{\Psi_{c}} (i \Slash{D} -M_{c} ) \Psi_{c} \biggr],
 \label{5DS}
\end{align}
where $M=0,1,2,3,y$, $D_{M} = \nabla_{M} + i A_{M} + i \hat{q}_{a} A^{(a)}_{M} + i \hat{q}_{c} A^{(c)}_{M}$ 
and $\hat{q}_{a}$ and $\hat{q}_{c}$ are the charges of $U(1)_{a}$ and $U(1)_{c}$ respectively.
5D Chern Simons terms associated with 4D Green-Schwarz mechanism
is neglected as already noted.
The field strengths are given by $F_{MN} = \partial_{M} A_{N} - \partial_{N} A_{M} + i [ A_{M} , A_{N} ]$ and $F^{(a,c)}_{MN} = \partial_{M} A^{(a,c)}_{N} - \partial_{N} A^{(a,c)}_{M}$ for the non-abelian gauge field and abelian gauge fields respectively.
The normalization of generator of $U(2)$ is chosen as $\tr (T_{a} T_{b}) = \delta_{ab}/2$,
hence the $U(2)$ gauge field is expanded as 
$A_M = \frac{{\1}_{2}}{2}A_M^{(0)} + \frac{\sigma^a}{2}A_M^{(a)}$,
where ${\1}_{2}$ is $2\times 2$ identity matrix and $\sigma^a$'s $(a=1,2,3)$ are the Pauli matrices.
Since the covariant derivative is acting on $\Psi_a$ as 
$D_M \Psi_a \ni i A_M \Psi_a= i  \frac{{\1}_{2}}{2}A_M^{(0)} \Psi_a + \cdots$,
$\Psi_a$ has $1/2$ charge against the overall $U(1)$. This is similar to $\Psi_c$, which has the opposite $U(1)$ charge.  
Here, $\Psi_{a,c} = (\psi_{a,c1}, \psi_{a,c2})^{\mathrm{T}}$ are doublets for the $SU(2)$
and $\psi_{a,c}$ are the 4D Dirac spinors.
The metric of $M_4 \times S^1 / \mathbb{Z}_2$
is written by $ds_{5}^2 = e^{-\sigma} g_{\mu\nu} d x^{\mu} d x^{\nu} + e^{2 \sigma} dy^2$, 
where $\sigma$ is the radion field, and gives 4D Einstein frame.
The size of $S^{1}/\mathbb{Z}_{2}$ is assumed to be $\pi L$ and we take $\langle \sigma \rangle = 0$
without loss of generality.
The graviphoton $g_{\mu y}$ is dropped since it is parity odd while
the 4D graviton $g_{\mu \nu}$ remains massless.
Then, the massive graviphoton mediates
the short-range force among particles which have the Kaluza-Klein (KK) charges,
and does not contribute to the SWGC condition.
On top of the usual periodic boundary condition of $S^1$,
the orbifold boundary conditon is given by 
\begin{align}
P A_{M} (x, -y) P^{-1} &= \eta_A A_M (x,y), 
\quad
A_M^{(a,c)} (x,-y)  = \eta_A A_M^{(a,c)} (x,y), \\
P \Psi_{a,d}(x,-y) &= \gamma_5 \Psi_{a,d}(x,y), 
\end{align}
where $P = \mathrm{diag}(+1, -1) \in U(2)$, $\eta_A = 1$ for $M=\mu=0,1,2,3$ and $\eta_A=-1$ for $M=y$.
Thus 4D massless modes read
\begin{align}
 &
 A_{\mu} = \left( \begin{array}{cc}
 A^{(b)}_{\mu} & \\
  &A^{(d)}_{\mu}
 \end{array}\right),
 \quad
 A_{y} = \left( \begin{array}{cc}
  & i \varphi / \sqrt{2} \\
 - i \varphi^\dag/\sqrt{2} & 
 \end{array}\right),
 \quad
 A^{(a)}_{\mu},
 \quad
 A^{(c)}_{\mu},
 \label{eq:orbifoldA}
 \\
 &
 \psi_{a1 R} ,~
 \psi_{a2 L},
 \quad
 \psi_{c1 R},~
 \psi_{c2 L}.
 \label{eq:orbifoldpsi}
\end{align}
where $A^{(b)}_{\mu}  := \frac{1}{2}(A_\mu^{(0)} + A_\mu^{(3)})$, 
$A^{(d)}_{\mu}  := \frac{1}{2}(A_\mu^{(0)} - A_\mu^{(3)})$,
$ \varphi = - (i A^{(1)}_{y} + A^{(2)}_{y} )/\sqrt{2}$ is the complex scalar originating 
from the $W$-boson of $y$-direction\footnote{
Here we define the complex scalar by multiplying the ordinary $W^{+}$-boson by $-i$ 
so that the effective Lagrangian of the zero modes reproduces Eq.~(\ref{eq:4-nodes}) after this orbifold projection.
},
and the $\psi_L$ ($\psi_R$) is the left-handed (right-handed) chiral fermion in 4D.
It turns out that there exists the gauge symmetry of $U(1)_a \times U(1)_b \times U(1)_c \times U(1)_d$.
As seen from the zero mode basis in the $U(2)$ gauge bosons, 
we find matter charges for the gauge symmetry: As for $(U(1)_a, U(1)_b, U(1)_c, U(1)_d)$,
$ \psi_{a1 R}^C \equiv \psi_{ab}:(+1,-1,0,0)$, $\psi_{c1 R}^C \equiv \psi_{bc}:(0,+1,-1,0)$, 
$\psi_{c2 L} \equiv \psi_{cd}:(0,0,1,-1)$, $\psi_{a2 L} \equiv \psi_{da} :(-1,0,0,1)$
and $\varphi: (0,1,0,-1)$.
This is the same field content as in the previous subsection, 
hence $\varphi$ is a neutral scalar under anomaly-free $U(1)$'s and will not be considered for the SWGC.
The scalar potential will be neglected as previously noted 
since the scalar potential including radion will depend on the model and the radion stabilization. 
Deriving the scalar potential and checking the strong SWGC for this is left for future work.
The 4D parameters are given by the 5D parameters with an assumption of $\langle \sigma \rangle = 0$.
For the details, see Appendix~\ref{appC}. 
The 4D Planck mass is associated with the 5D gravitational coupling $\kappa_{5}$ as
\begin{align}
 M_{\mathrm{Pl}}^2 = \frac{\pi L }{\kappa_{5}^2},
\end{align}
and the gauge couplings in 4D are expressed by 
\begin{align}
\frac{2}{g^2_{2}} :=  \frac{1}{g_b^2} = \frac{1}{g_d^2}=  \frac{2 \pi L }{\hat{g}_{2}^2} ,
 \qquad
 \frac{1}{g_{a}^2} := \frac{ \pi L }{\hat{g}_{a}^2},
 \quad
 \frac{1}{g_{c}^2} := \frac{ \pi L }{\hat{g}_{c}^2}.
 \label{eq:gaugecoupling}
\end{align}
This is because we have the gauge kinetic term ${\cal L} = -2/4g_2^2 \bigl(F_{\mu \nu}^{(b)}\bigr)^2 
-2/4g_2^2 \bigl(F_{\mu \nu}^{(d)}\bigr)^2$ via the symmetry breaking of  $U(2) \to U(1)_{b} \times U(1)_{d}$.
With these, the anomaly-free couplings are defined as previously:
\begin{align}
 \frac{1}{e_{X}^2} =& \frac{c^2}{g_{a}^2} + \frac{2}{g_{2}^2} + \frac{c^2}{g_{c}^2} + \frac{2}{g_{2}^2},
 \label{eq:g_{X}^2SWGC}
 \\
 \frac{1}{e_{X'}^2} = & \frac{1/c^2}{g_{a}^2} + \frac{2}{g_{2}^2} + \frac{1/c^2}{g_{c}^2} + \frac{2}{g_{2}^2},
 \label{eq:g_{X'}^2SWGC}
\end{align}
where a free parameter $c$ is a rational number.
The Yukawa couplings between $\varphi$ and  $\psi$'s are given by%
\begin{align}
 y = y' =
 \frac{ \hat{g}_{2}}{\sqrt{ 2\pi L}}
  = \frac{g_{2}}{\sqrt{2}}.
 \label{eq:yukawagauge}
\end{align}
Here, $y$ and $y'$ are the same definition as in the previous subsection.
This equation relates the Yukawa coupling to the gauge coupling.
In the presence of a light $\mathrm{Re} (\varphi )$ and the radion, the SWGC inequality for zero mode fermions
reads
\begin{align}
 ( 1- c)^2 e_{X}^2 + \biggl( 1 + \frac{1}{c} \biggr)^2 e_{X'}^2 \geq \frac{y^2}{2} 
  + \left(  \frac{1}{2} + \frac{1}{6} \right) \frac{M^2}{M_{\mathrm{Pl}}^2}  
 \label{eq:SWGCapp0} 
\end{align}
where $M$ has the same definition as in the previous subsection\footnote{
We have factored out the common radion dependence $e^{-\sigma}$, 
}
and $1/6$ comes from the radion exchange via 
$ye^{-\sigma_c/\sqrt{6}M_{\rm Pl}} {\rm Re}(\varphi) \bar\psi \psi$ with the canonically normalized radion 
$\sigma_c = \sqrt{3/2}M_{\rm Pl} \sigma$.
We have neglected momentum-dependent terms induced by the radion exchange
with terms of $\bar\psi \gamma^\mu \psi \partial_\mu \sigma $.
If $ \mathrm{Re} (\varphi)$ is sufficiently heavy,
the yukawa interaction in this equation can be neglected and the WGC can be then easily satisfied.
Gravitational interactions including radion exchange will be numerically neglected below 
as in the previous subsection owing to the Planck-suppressed interaction within the effective field theory.
For $M/g_2 M_{\rm Pl} \lesssim 0.1$ and $p/g_2 M_{\rm Pl} \lesssim 0.1$, 
where $p$ is the momentum of a test fermion, there are not significant differences compared to the plots shown below.
Substituting the above couplings given by Eqs.~(\ref{eq:g_{X}^2SWGC})--(\ref{eq:yukawagauge}) to Eq.~(\ref{eq:SWGCapp0}), we then find the SWGC condition
\begin{align}
 \frac{(1-c)^2}{(c^2/2) ( g_2^2/ g_{a}^2 + g_2^2/g_{c}^2) +2 } 
 + \frac{ (1 + 1/c)^2}{(1/2c^2) (g_2^2 /g_{a}^2 +g_2^2 /g_{c}^2) +2 }
 \gtrsim
 \frac{1}{4}.
 \label{eq:SWGCapp}
\end{align}
This is also obtained when the parameters in Eq.~(\ref{eq:exSWGC}) are replaced
as $g^2 \to g^2_2/2$ and $(Y/g)^2 \to 1/2$.
This gives a constraint between $c$ and $X := g_2^2/ g_{a}^2 + g_2^2/g_{c}^2$. 

As for the KK modes or the massive parity odd ones, 
a similar equation to Eq.~(\ref{eq:SWGCapp0}) will be hold.
It is noted that massive gauge bosons do not contribute to long-range forces and
all bosons including scalar zero mode are neutral under anomaly-free $U(1)$'s and fermions 
with non-trivial gauge charges are considered for the SWGC. 
Parity odd fermions of $\psi_{a1L},~\psi_{a2R}, \psi_{c1L}$ and $\psi_{a2R}$ have
opposite charges to zero mode fermions.
KK modes of a field have the same charge as that of 
the lightest mode.
Yukawa couplings that are invariant under ${\mathbb Z}_2$ projection are given by 
$\phi_{\rm even}\psi_{\rm even}\psi_{\rm even}$,~$\phi_{\rm even}\psi_{\rm odd}\psi_{\rm odd}$,
~$\phi_{\rm odd}\psi_{\rm odd}\psi_{\rm even}$,
where $\phi_{\rm even}$ ($\phi_{\rm odd}$) is an even (odd) parity scalar
and $\psi_{\rm even}$ ($\psi_{\rm odd}$) is an even (odd) parity fermion.
As massive scalars do not contribute to a long-range force,
Yukawa couplings relevant to the SWGC are associated with $\varphi$:
$\varphi \psi_{\rm even}\psi_{\rm even},~\varphi \psi_{\rm odd}\psi_{\rm odd}$.
After integration over the extra dimension, 
we will find Yukawa couplings of
$y \varphi \psi_{n}'\psi_{n}$ in addition to KK mass terms 
$ (n/L )\bar{\psi}_{n}\psi_{n}+(n/L )\bar{\psi'}_{n}\psi_{n}'$ for $n$-th KK modes
with the canonically normalized kinetic terms 
(up to the radion dependence).
Then $n$-th KK mass eigenstates will have mass $M^2 = (n/L \pm y{\rm Re}(\varphi))^2$.
As the SWGC could be violated by heavy KK modes,
it is necessary to check whether lighter modes including the zero modes satisfy the SWGC.

%
\begin{figure}[t]
\centering
\includegraphics[scale=0.3]{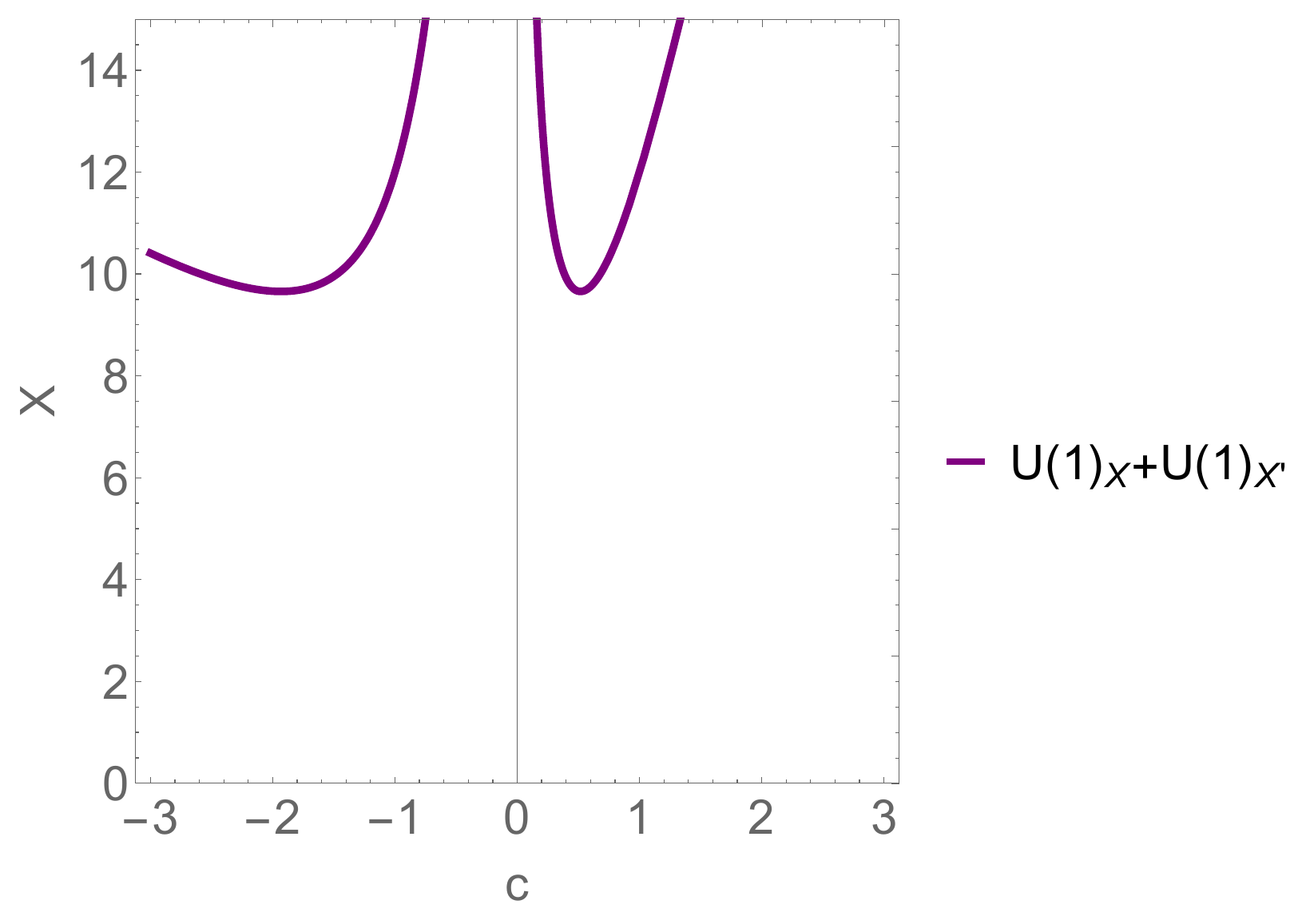}
\includegraphics[scale=0.3]{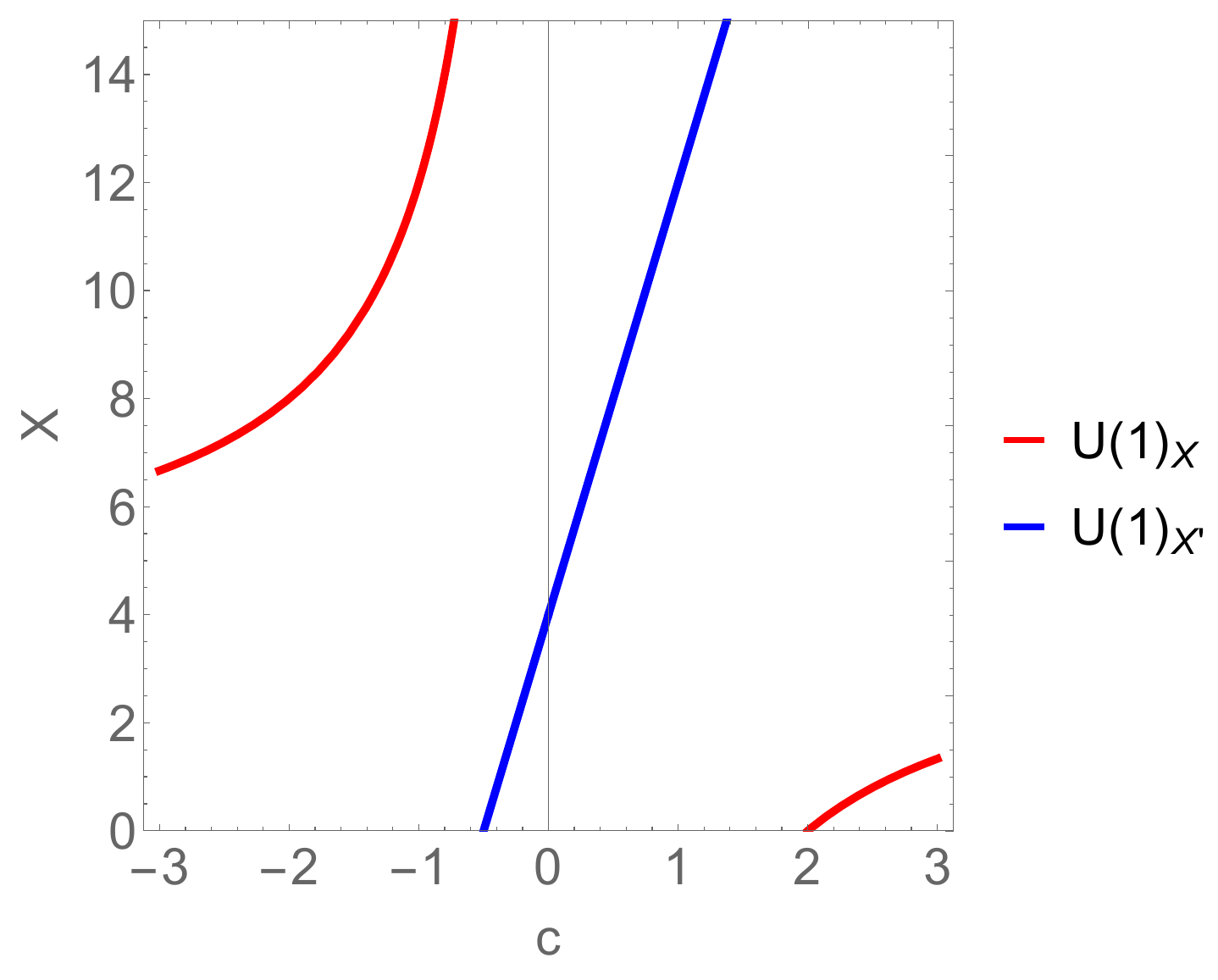}
\caption{
The left panel: 
Plots of the SWGC conditons (\ref{eq:SWGCapp}) in the $(c, X)$-planes.
The condition is saturated on each lines, below which
there exist an allowed region.
In the presence of mass, the SWGC is violated on each line.
The right panel: A similar plot with only $U(1)_X$ (red) and one with only $ U(1)_{X'}$ (blue).
\label{fig:resSWGC}
}
\end{figure}
%

%
Fig.~\ref{fig:resSWGC} shows the plots of the SWGC (\ref{eq:SWGCapp}) in the $(c, X)$ plane. 
The each line saturates Eq.~(\ref{eq:SWGCapp}), hence the allowed region exists below them.
In the presence of mass, the SWGC is violated on each line.
The these plots are symmetric under $c \to -1/c$.
The left panel shows the constraint when there exists two anomaly-free $U(1)$'s.
This is very similar to the top-right one of Fig.~\ref{fig:testplot} for a small value of Yukawa.
For a large $X$, not only $U(1)_b \times U(1)_d$ gauge coupling but also
Yukawa coupling $\propto g_2$ become much stronger than $g_a$ or $g_c$,
and hence there exist an upper bound on $X$.
It is noted that in the context of the string theory a large $X$ may imply 
a big discrepancy among moduli vacuum expectation values. 
In the vincity of $c=\pm 1$, matter becomes neutral against either one of the anomaly-free $U(1)$'s,
then the constraint becomes tighter.
In the right panel, plots show the SWGC constraint with $e_{X} = 0$ or $e_{X'} = 0$,
when only the gauge boson of either $U(1)_X$ or $U(1)_{X'}$ remains massless 
owing to a St\"uckelberg coupling as often seen in concrete string models.
%

\section{Conclusion}
\label{sec:conclusion}
We have studied the (S)WGC in several types of quiver gauge theories 
with $U(1)^{k}$ gauge symmetry in the presence of
bi-fundamental chiral fermions leading to the chiral anomalies, which is supposed to be canceled by the 
Green-Schwarz mechanism.
The theories which we consider possesses a cyclic ${\mathbb Z}_k$ symmetry
associated with a shift of the label of the gauge groups.
As a consequence of this, we can study anomalies in the models systematically 
and the (S)WGC constraints on the gauge couplings.
We identified concretely the anomaly-free $U(1)$'s and their gauge couplings
obtained via linear combinations of the original $U(1)$'s.
Then, ${\mathbb Z}_k$ symmetry can be broken in general.
In the large $k$ limit, an anomaly-free gauge coupling becomes very weak as $e \sim k^{-1/2}$,
and there exists an upper bound on $k$ if the WGC is correct 
and the mass for a test particle remains in the large $k$ limit.
This may be regarded as an example of the weak coupling conjecture.
For quiver theories with $U(1)^{2k-1}$, 
an unique anomaly-free $U(1)$ is proportional to $\sum_{i=1}^{2k-1} U(1)_{i}$ and
all matters are neutral under the anomaly-free $U(1)$.
There exist charged matters in the presence of vector-like pairs, and $\mathbb{Z}_{2k-1}$ symmetry is broken then.
For quiver theories with $U(1)^{2k}$ gauge symmetry, there exist two anomaly-free $U(1)$'s 
and charged matters under these gauge groups, and $\mathbb{Z}_{2k}$ symmetry is broken in general. 
Even if the gauge couplings of the the anomaly-free $U(1)$'s receive quantum corrections,
the IR couplings will remain very weak in the large $k$ limit
since $U(1)$ couplings are generally asymptotic non-free as far as the pertubation theory is valid. 
We have numerically studied also the SWGC in $U(1)^4$ theory in the presence of a complex scalar field,
and construct a similar model based on a 5D orbifold.
It turns out that a much larger Yukawa coupling than gauge couplings is forbidden 
and also that a big discrepancy among gauge couplings is disfavored.
A special linear combination for realizing the anomaly-free $U(1)$'s can be also be disfavored,
since matter charge becomes small then.

So far, we neglected kinetic mixings $\chi_{ij} F_{\mu \nu}^i F^{j \mu \nu}$ among gauge fields.
If we have such terms, we may have a kinetic mixing of $\chi F_{\mu \nu}^X F^{X' \mu \nu}$,
where $\chi \sim \sum_{i,j}^k \chi_{ij}c_i c_j' $,
for anomaly-free $U(1)_X = \sum_i c_i U(1)_i$ and $U(1)_{X'}=\sum_i c_i' U(1)_i$ with $c_i$'s $={\cal O}(1)$. 
If the mixing $\chi$ is at most of ${\cal O}(k^{3/2 - \alpha})~(\alpha > 0)$ in the large $k$ limit, 
the WGC can still be violated as in Sec.~\ref{Sec:quiver}.
This is because the canonically normalized mixing is given by $e_X e_{X'}\chi $ 
and hence an induced coupling of a fermion to an anomay-free gauge field
is proportional to $e_X^2 e_{X'}\chi$ or $e_{X}e_{X'}^2 \chi$ that are scaling as $k^{-\alpha}$ then.
However, if $\chi = {\cal O}(k^2)$ in the large $k$ limit, the WGC can be satisfied.

In the Sec.~\ref{Sec:4node}, the scalar $\varphi$ is a singlet under the anomaly-free gauge groups,
and we did not discuss the detail of the scalar potential in addition to the radion.
Hence it may be an interesting challenge to check the strong SWGC within a fixed model. 
This is left for future work.

It will be worth to investigate the (S)WGC in theories with more general gauge groups.
In actual string compactifications, the number of closed string axions
is known to be finite and depends on the Hodge number of compactification manifold.
Some of the axions play an important role to cancel anomalies through the Green-Schwarz mechanism.
Hence, the number of anomalous $U(1)$ gauge theories,
which is $k-1$ or $k-2$ in our cases, should be constrained by the number of such axions. 
If the anomalies are independent among the anomalous theories,
the number of anomalous $U(1)$'s can be less than that of axions for anomaly cancellation.
Also in the string theory, the conjecture would constrain brane configuration and moduli values.
If one starts with 10D super Yang-Mills theory, 4D effective action including an anomaly-free $U(1)$ may be given by
\cite{Cremades:2003qj, Cremades:2004wa}
\begin{align}
 \cal{L} = -c \frac{S}{4} (F_{\mu \nu})^2 - \frac{\vartheta(\tau)}{\sqrt{S}} \phi \bar{\psi} \psi + \cdots,
\end{align}
where $S$ is the 4D dilaton, $\tau$ is a complex structure modulus, 
and a rational number $c$ originates from a linear combination of $U(1)$'s
depends on 
brane configuration of the number of branes and magnetic fluxes.
The SWGC of $e^2 \gtrsim y^2$ (up to mass term) for matter fermion may read
\begin{align}
c \lesssim \frac{1}{|\vartheta(\tau)|^2}.  
\end{align}
However, it will be required a deep understanding of the string theory 
or concrete effective field theories including (non-abelian) Dirac-Born-Infeld action
to study the SWGC constraints on moduli space for consistent gauge theories 
in the presence of the Green-Schwarz mechanism.
This is also left for future works.
%


\section*{Acknowledgments}
\noindent
The authors thank Toshifumi Noumi, Tatsuo Kobayashi, Takahiro Terada and Yuta Hamada
for useful discussions and comments.
The authors thank also Yoshiyuki Tatsuta at the early stage of collaboration.
The work of Y.A. is supported by JSPS Grant-in-Aid for Scientific Research, No.~JP20J11901.
\appendix

\section{Anomalies in string-inspired (SUSY) gauge theories}
\label{appA}

We discuss anomaly-free $U(1)$'s in $U(N)^{k}$ quiver gauge theories 
inspired by the string theory.
We focus only on certain types of quiver theories 
considered in Sec.~\ref{Sec:quiver} in this section.
Hereafter, $N_{a}$ denotes the rank of the gauge group of $U(N_a)$ at the $a$-node,
and $n_{ab}$ shows the number of bi-fundamental matter fields which correspond to
that of arrows connecting between $a$-node and $b$-node in the quiver diagram.
%

\subsection{$U(N)^3$}
%
\begin{figure}[t]
\centering
\includegraphics[scale=0.3]{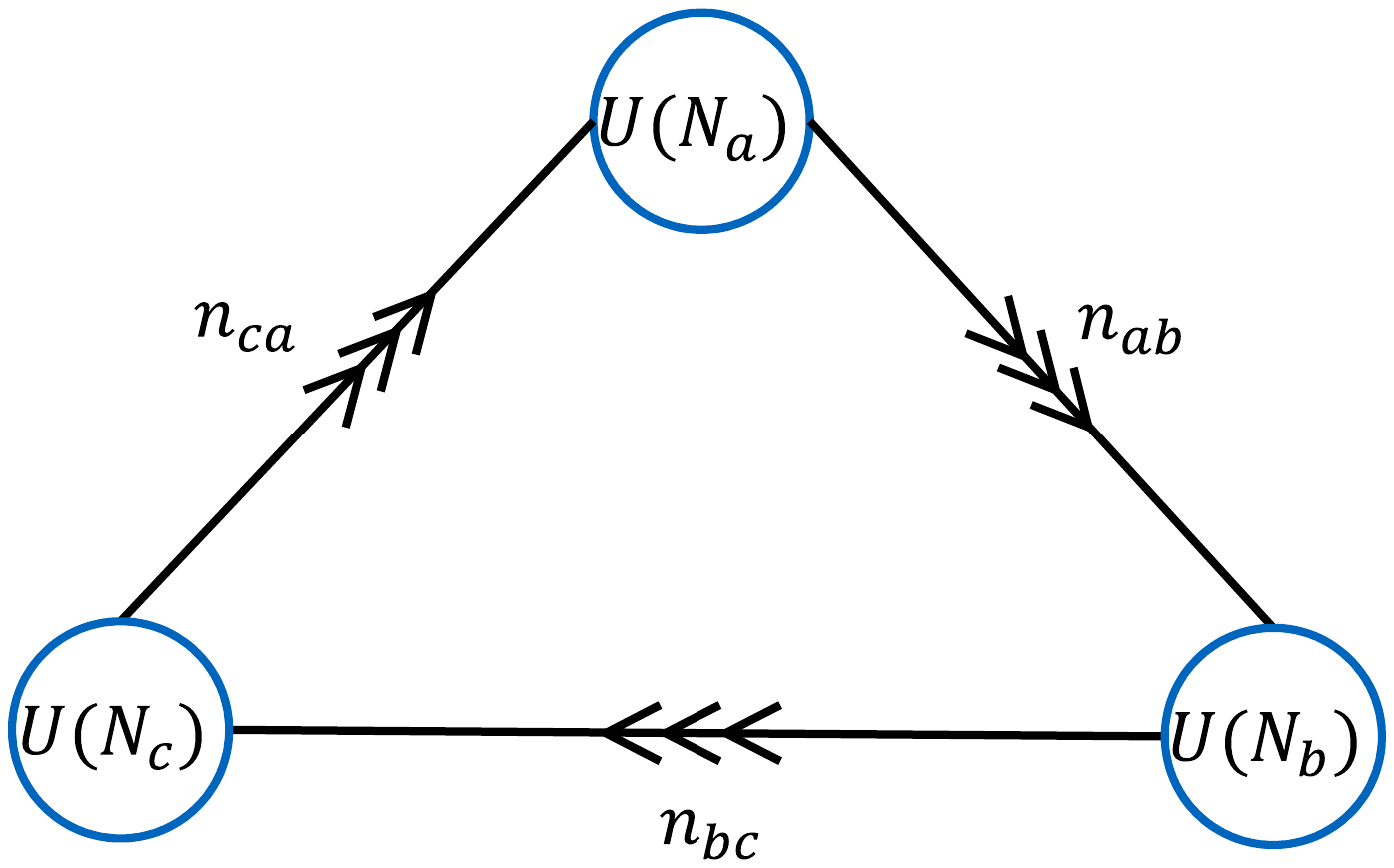}
\caption{
A quiver diagram of $U(N)^3$ gauge theory.
}
\label{fig:quiverN3}
\end{figure}
%

%
We consider a $U(N)^3$ quiver gauge theory as shown in Fig.~\ref{fig:quiverN3}, 
and identify an anomaly-free $U(1)$.
To this end, we calculate chiral anomalies and mixed anomalies.
Then, we find the constraints on the ranks of gauge groups and the numbers of generations.
For a consistent theory, the non-abelian cubic anomalies $\mathcal{A}_{SU(N_{a,b,c})^3}$ give the following constraints,
\begin{align}
 \mathcal{A}_{SU( N_{a} )^{3}} \propto & ( n_{ab} N_{b} - n_{ca} N_{c} ) \equiv 0,
 \\
 \mathcal{A}_{SU (N_{b})^{3}} \propto & ( n_{bc} N_{c} - n_{ab} N_{a} ) \equiv 0,
 \\
 \mathcal{A}_{SU (N_{c})^{3}} \propto & ( n_{ca} N_{a} - n_{bc} N_{b} ) \equiv 0.
\end{align}
With these equations, the ranks of the gauge groups are related as
\begin{align}
 N_{a} = \frac{n_{bc}}{n_{ab}} N_{c} \in \mathbb{N},
 \quad
 N_{b} = \frac{ n_{ca}}{n_{ab}} N_{c} \in \mathbb{N}.
 \label{eq:solquiver3-1}
\end{align}
Thus, we find $N_{a} = N_{b} = N_{c}$ for $n_{ab}=n_{bc}=n_{ca}$.
The divergences of the chiral currents $j^{a,b,c}$ for $U(1)_{a,b,c}$ are given by
\begin{align}
 \partial \cdot j^{a} = & N_{a} ( N_{b} n_{ab} Q_{b} - N_{c} n_{ac} Q_{c} ) 
 + N_{a} ( n_{ab} N_{b} - n_{ca} N_{c} )Q_a  + \mathcal{A}_{U(1)_{a} G^2} ,
 \\
 \partial \cdot j^{b} = & N_{b} ( N_{c} n_{bc} Q_{c} - N_{a} n_{ab} Q_{a} )
  +  N_{b} ( n_{bc} N_{c} - n_{ab} N_{a} )Q_b +\mathcal{A}_{U(1)_{b} G^2} ,
 \\
 \partial \cdot j^{c} = & N_{c} ( N_{a} n_{ca} Q_{a} - N_{b} n_{bc} Q_{b} ) 
 + N_{c} (n_{ca} N_{a} - n_{bc} N_{b} )Q_c + \mathcal{A}_{U(1)_{c} G^2}  ,
\end{align}
where $Q_{x} = Q^{U(1)_x} + \frac{1}{N_x}Q^{SU(N_x)}$,
$Q^{G_x} = \frac{1}{16\pi^2} \epsilon^{\mu\nu\rho\sigma} \mathrm{tr}(F^{(x)}_{\mu\nu} F^{(x)}_{\rho\sigma})$
for $x=a,b,c$,
and tr$(T^i T^j)=\delta^{ij}/2$ for $U(N)$ generators $T^i$'s.
These include anomalies of $U(1)^3$, $U(1)SU(N)^2$ and
the mixed anomalies between the gravity and $U(1)$'s, which are denoted by
$\mathcal{A}_{U(1)_{a,b,c} G^2}$.
We impose that they are vanishing:
\begin{align}
 \mathcal{A}_{U(1)_{a} G^2} \propto N_{a} ( n_{ab} N_{b} - n_{ca} N_{c} ) \equiv 0,
 \\
 \mathcal{A}_{U(1)_{b} G^2} \propto N_{b} ( n_{bc} N_{c} - n_{ab} N_{a} ) \equiv 0,
 \\
 \mathcal{A}_{U(1)_{c} G^2} \propto N_{c} (n_{ca} N_{a} - n_{bc} N_{b} ) \equiv0.
\end{align}
This is the same condition as in the non-abelian anomalies.
There are not exist charged fields under all $(U(1)_{a}, U(1)_{b}, U(1)_{c})$, 
then the anomaly between $U(1)_{a} U(1)_{b} U(1)_{c}$ vanishes automatically.
Then we find
\begin{align}
 \partial \cdot j^{a} \equiv & N_{a} ( N_{b} n_{ab} Q_{b} - N_{c} n_{ac} Q_{c} ) ,
 \\
 \partial \cdot j^{b} \equiv& N_{b} ( N_{c} n_{bc} Q_{c} - N_{a} n_{ab} Q_{a} ) ,
 \\
 \partial \cdot j^{c} \equiv & N_{c} ( N_{a} n_{ca} Q_{a} - N_{b} n_{bc} Q_{b} ) .
\end{align}
To identify the anomaly-free $U(1)$ we define it as
\begin{align}
 U (1)_{X} : = \frac{c_{a}}{N_{a}} U(1)_{a} + \frac{c_{b}}{N_{b}} U(1)_{b} + \frac{c_{c}}{N_{c}} U(1)_{c},
\end{align}
and we impose that the divergence of the current associated with $U(1)_{X}$ vanishes
\begin{align}
 \partial \cdot j^{X} = & \sum_{x = a,b,c} \frac{c_{x}}{N_{x}} \partial \cdot j^x
 \nonumber\\
 =& N_{a} ( c_{c} n_{ca} - c_{b} n_{ab} ) Q_{a} + N_{b} (c_{a} n_{ab} - c_{c} n_{bc} ) Q_{b} + N_{c} ( c_{b} n_{bc} - c_{a} n_{ca} ) Q_{c}
 \equiv 0.
\end{align}
From this equation, the coefficients satisfy the following conditions
\begin{align}
 c_{a} = \frac{n_{bc}}{n_{ab}} c_{c},
 \quad
 c_{b} = \frac{n_{ca}}{n_{ab}} c_{c}.
 \label{eq:solquiver3-2}
\end{align}
We take $c_{c}=1$ and use Eqs.~(\ref{eq:solquiver3-1}) and (\ref{eq:solquiver3-2}), 
then the anomaly-free $U(1)$ is given by
\begin{align}
 U(1)_{X} = \frac{c_{c}}{N_{c}} \bigl( U(1)_{a} + U(1)_{b} + U(1)_{c} \bigr).
\end{align}
It is noted that all fields are neutral matter under this anomaly-free $U(1)$.
The anomaly-free gauge coupling is given by
\begin{align}
 \frac{1}{e_{X}^2} = \frac{1}{g_{a}^2} + \frac{1}{g_{b}^2} +\frac{1}{g_{c}^2}.
\end{align}
%

\subsection{$U(N)^4$}
%
\begin{figure}[t]
\centering
\includegraphics[scale=0.35]{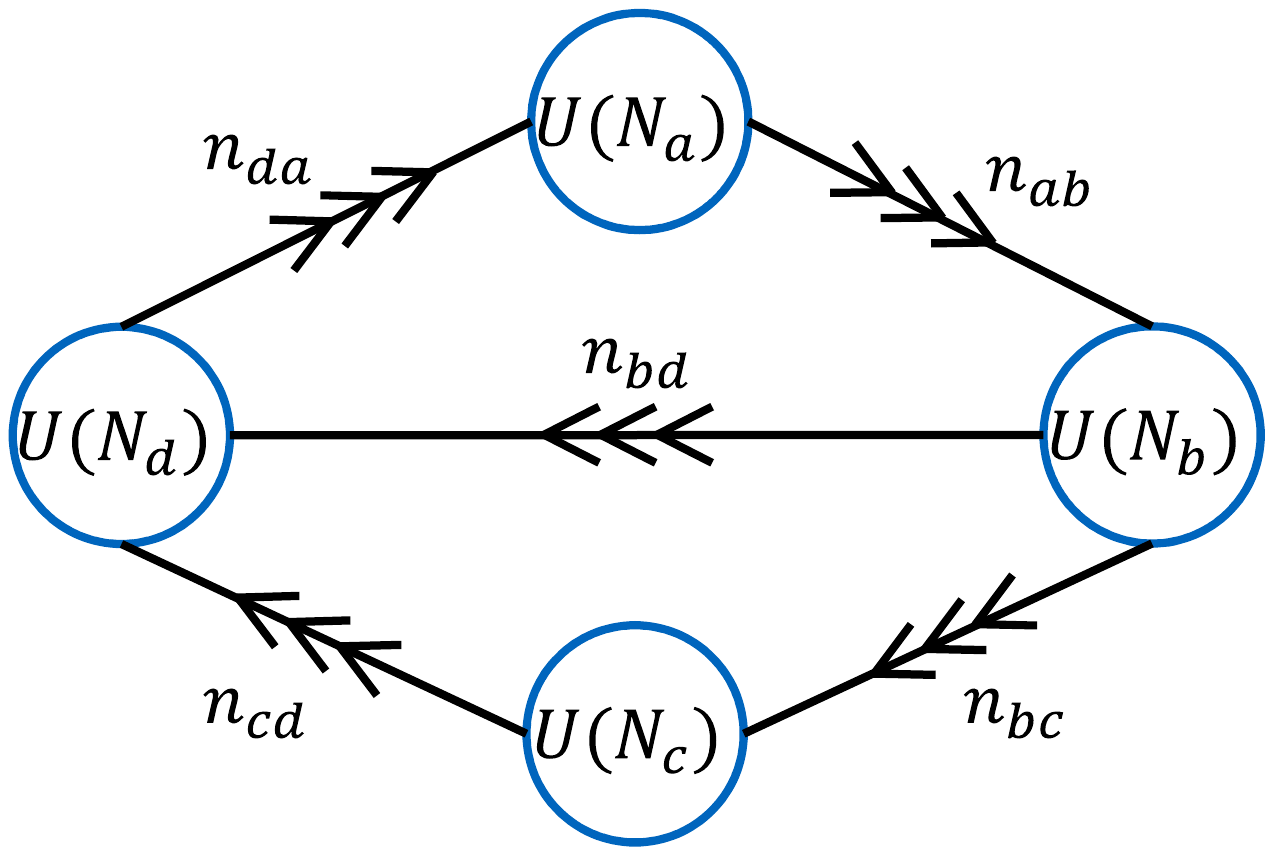}
\caption{
$U(N)^4$ quiver diagram.
}
\label{fig:quiverN4ex2-2}
\end{figure}
%

%
In this case, we impose that the anomaly coefficients of non-abelian cubic anomaly are vanishing:
\begin{align}
 \mathcal{A}_{SU(N_{a})^3} \propto & ( n_{ab} N_{b} - n_{da} N_{d} ) \equiv 0,
 \\
 \mathcal{A}_{SU(N_{b})^3} \propto & ( - n_{ab} N_{a} + n_{bc} N_{c} + n_{bd} N_{d} ) \equiv 0,
 \\
 \mathcal{A}_{SU(N_{c})^3} \propto & ( - n_{bc} N_{b} + n_{cd} N_{d} ) \equiv0,
 \\
 \mathcal{A}_{SU(N_{d})^3} \propto & ( n_{da} N_{a} - n_{bd} N_{b} - n_{cd} N_{c} ) \equiv0.
\end{align}
Solving these equations, 
we find that the ranks of gauge groups and the numbers of generations have the following relations, 
\begin{align}
 N_{b} = \frac{n_{da}}{n_{ab}} N_{d} \in \mathbb{N},
 \quad
 N_{c} = \frac{n_{da}}{n_{cd}} \biggl[ N_{a} - \biggl( \frac{n_{bd}}{n_{ab}} \biggr) N_{d} \biggr] \in \mathbb{N},
 \quad
 \frac{n_{da}}{n_{ab}} = \frac{n_{cd}}{n_{bc}}.
 \label{eq:solap1}
\end{align}
The cancellation of the mixed anomaly between the gravity and $U(1)$'s 
imposes the same constraints as above: 
\begin{align}
 \mathcal{A}_{U(1)_{a} G^{2}} \propto &  N_{a} ( n_{ab} N_{b} -n_{da} N_{d} ) \equiv 0,
 \\
 \mathcal{A}_{U(1)_{b} G^{2}} \propto & N_{b} ( -n_{ab} N_{a} + n_{bc} N_{c} + n_{bd} N_{d}) \equiv 0,
 \\
 \mathcal{A}_{U(1)_{c} G^{2}} \propto & N_{c} ( -n_{bc} N_{b} + n_{cd} N_{d} ) \equiv 0,
 \\
 \mathcal{A}_{U(1)_{d} G^{2}} \propto & N_{d} ( n_{da} N_{a} - n_{bd} N_{b} - n_{cd} N_{c} ) \equiv 0.
\end{align}
The divergences of the $U(1)$ currents are expressed as
\begin{align}
 \partial \cdot j^{a} \equiv & N_{a} ( N_{b} n_{ab} Q_{b} - N_{d} n_{da} Q_{d}),
 \\
 \partial \cdot j^{b} \equiv & N_{b} ( - N_{a} n_{ab} Q_{a} + N_{b} n_{bc} Q_{c} + N_{d} n_{bd} Q_{d} ),
 \\
 \partial \cdot j^{c} \equiv & N_{c} ( - N_{b} n_{bc} Q_{b} + N_{d} n_{cd} Q_{d}),
 \\
 \partial \cdot j^{d} \equiv & N_{d} (  N_{a} n_{da} Q_{a} - N_{b} n_{bd} Q_{b} - N_{c} n_{cd} Q_{c}),
\end{align}
where we used vanishing conditions of non-abelian anomalies.
We define the anomaly-free $U(1)$ by the following equation 
as in the previous subsection
\begin{align}
 U(1)_{X} := \frac{c_{a}}{N_{a}} U(1)_{a} + \frac{c_{b}}{N_{b}} U(1)_{b} + \frac{c_{c}}{N_{c}} U(1)_{c} + \frac{c_{d}}{N_{d}} U(1)_{d},
\end{align}
and impose the current divergence associated with this $U(1)_{X}$ is vanishing
\begin{align}
 \sum_{x =  a, b, c, d} \frac{c_{x}}{N_{x}} \partial \cdot j^{x} =&
 N_{a} ( - c_{b} n_{ab} + n_{d} n_{da} ) Q_{a}
 +N_{b}( c_{a} n_{ab} - c_{c} n_{bc} - c_{d} n_{bd} ) Q_{b}
 \nonumber\\
 &+ N_{c} (c_{b} n_{bc} - c_{d} n_{cd} ) Q_{c}
  +N_{d} ( -c_{a} n_{da} + c_{b} n_{bd} + c_{c} n_{cd} ) Q_{d}
 \nonumber\\
 \equiv & 0.
\end{align}
Solving these equations for the coefficients $c_{x}$, we get the following relations
\begin{align}
 c_{b} = \frac{n_{da}}{n_{ab}} c_{d},
 \quad
 c_{c} = \frac{n_{da}}{n_{cd}} \biggl[ c_{a} - \biggl( \frac{n_{bd}}{n_{ab}} \biggr) c_{d} \biggr],
 \quad
 \frac{n_{da}}{n_{ab}} = \frac{n_{cd}}{n_{bc}}.
 \label{eq:solap2}
\end{align}
From Eqs.~(\ref{eq:solap1}) and (\ref{eq:solap2}), the coefficient $c_{a}$ (or $c_{b}$) is a free parameter.
In oder to solve these equations,
we shall impose some assumptions.
Here we will list some examples satisfying these equations.
\begin{itemize}
\item
$\forall~ N=1$
\begin{itemize}
\item
$n_{bd}=0$\\
A solution is
\begin{align}
 N_{a} = N_{b} = N_{c} = N_{d} = 1,
 \quad
 n_{ab} = n_{bc} = n_{cd} = n_{da},
 \quad
 n_{bd}=0.
\end{align}
This is similar to the quiver gauge theory shown in Fig.~\ref{fig:quiverN4simp1}.
The two independent anomaly-free $U(1)$'s are generally given by Eqs.~(\ref{eq:U14U1X}) and (\ref{eq:U14U1X'}),
and the corresponding gauge couplings are expressed as Eq.~(\ref{eq:U14gX}) and (\ref{eq:U14gX'}).
\item
$n_{bd}=2$\\
A solutions is given by
\begin{align}
 N_{a} = N_{b} = N_{c} = N_{d} =1,
 \quad
 n_{ab} = n_{da} = -n_{bc} =- n_{cd} = 1,
 \quad
 n_{bd}=2.
\end{align}
The minus sign represents the opposite arrow of Fig.~\ref{fig:quiverN4ex2-2}.
The independent anomaly-free $U(1)$'s are defined by
\begin{align}
 U(1)_{X} =& c U(1)_{a} + U(1)_{b} + (2-c) U(1)_{c} + U(1)_{d},
 \\
 U(1)_{X'} = & \frac{c-3}{c-1} U(1)_{a} + U(1)_{b} + \frac{c+1}{c-1} U(1)_{c} + U(1)_{d}.
\end{align}
For these anomaly-free $U(1)$'s, $bd$ matters are neutral.
The anomaly-free gauge couplings are given by
\begin{align}
 \frac{1}{e_{X}^2} =& \frac{c^2}{g_{a}^2} + \frac{1}{g_{b}^2} + \frac{(2-c)^2}{g_{c}^2} + \frac{1}{g_{d}^2},
 \\
 \frac{1}{e_{X'}^2} =& \biggl( \frac{c-3}{c-1} \biggr)^2 \frac{1}{g_{a}^2} + \frac{1}{g_{b}^2} + \biggl( \frac{c+1}{c-1} \biggr)^2 \frac{1}{g_{c}^2} + \frac{1}{g_{d}^2}.
\end{align}
\end{itemize}
\item
$\forall~|n|=1$\\
A solution is given by
\begin{align}
 N_{b} = N_{d} =2,
 \quad
 N_{a} = N_{c}=1,
 \quad
 n_{ab}= n_{da} = n_{bd} = -n_{bc} = -n_{cd}=1.
\end{align}
The independent anomaly-free $U(1)$'s for this solution is defined as
\begin{align}
 U(1)_{X} =& c U(1)_{a} + \frac{1}{2} U(1)_{b} + (1-c) U(1)_{c} + \frac{1}{2} U(1)_{d},
 \\
 U(1)_{X'} = & \frac{2c-3}{4c-2} U(1)_{a} + \frac{1}{2} U(1)_{b} + \frac{2 c + 1}{4c-2} U(1)_{c} + \frac{1}{2} U(1)_{d}.
\end{align}
$bd$ matter is neutral for these anomaly-free gauge groups.
The gauge couplings are given by
\begin{align}
 \frac{1}{e_{X}^2} = & \frac{c^2}{g_{a}^2} \frac{1/2}{g_{b}^2} + \frac{(1-c)^2}{g_{c}^2} + \frac{1/2}{g_{d}^2},
 \\
 \frac{1}{e_{X'}^2} =& \biggl( \frac{2c -3}{4c -2 } \biggr)^2 \frac{1}{g_{a}^2} 
 + \frac{1/2}{g_{b}^2} + \biggl( \frac{ 2c+ 1}{4c-2} \biggr)^2 \frac{1}{g_{c}^2}
 + \frac{1/2}{g_{d}^2}.
\end{align}
It is noted that a coefficient of $1/g_{b,d}^2$ is given by $1/2= N_{b,d}\cdot (1/2)^2$ for $N_{b,d}=2$.
\end{itemize}

\section{Models inspired by the SM}
\label{Sec:appB}

We consider two quiver models with $U(1)^4$ and $U(1)^5$ symmetries inspired by the SM.
These are different from the models exhibited in the Sec.~\ref{Sec:quiver} in terms of chiral fermions.
We show just that the anomaly-free gauge couplings are still given by a linear combination of the original couplings. 
The SM might not originate from a gauge symmetry that has too many $U(1)$'s.

\subsection{A model inspired by Pati-Salam}
%
\begin{figure}[t]
\centering
\includegraphics[scale=0.26]{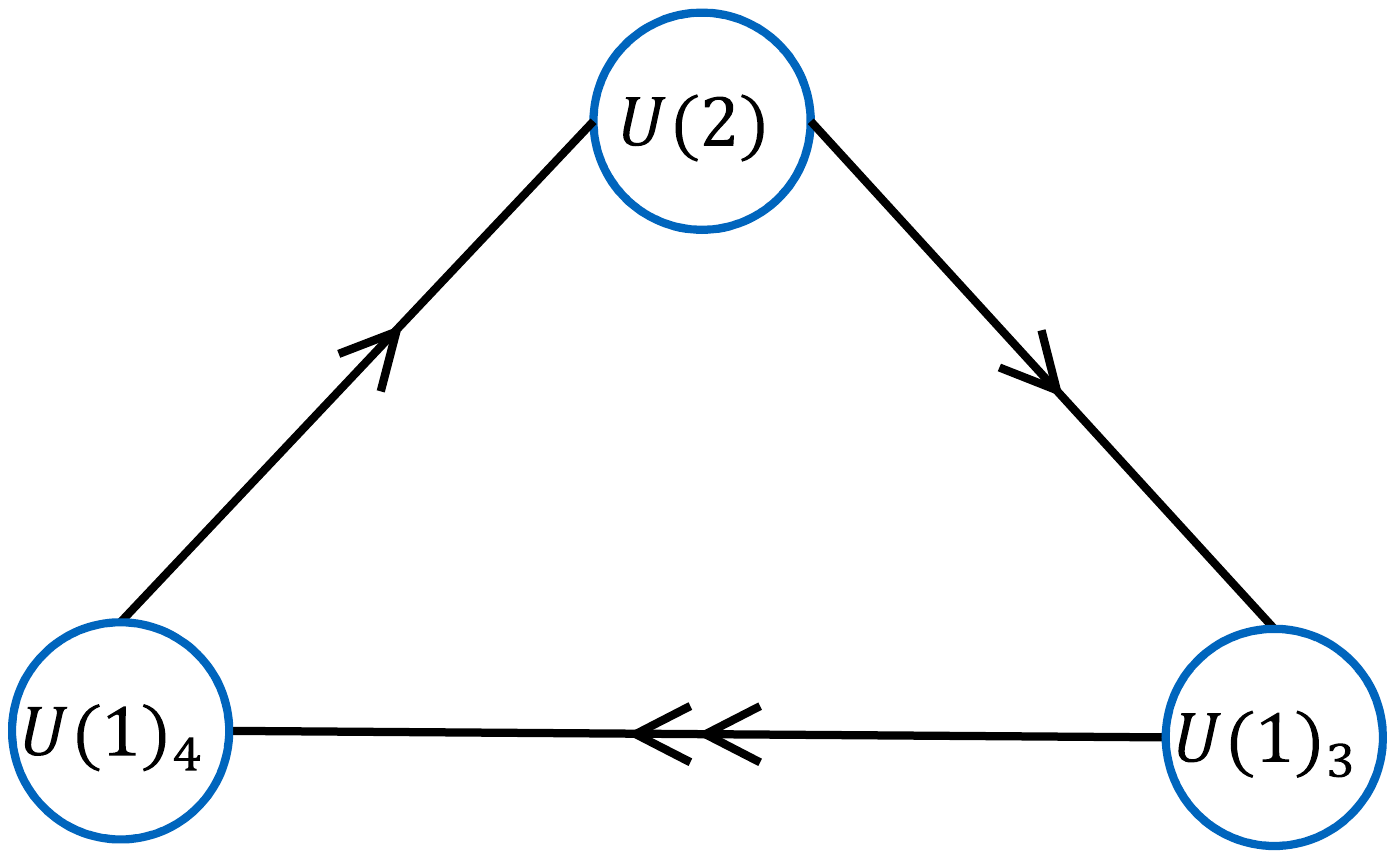}
\includegraphics[scale=0.26]{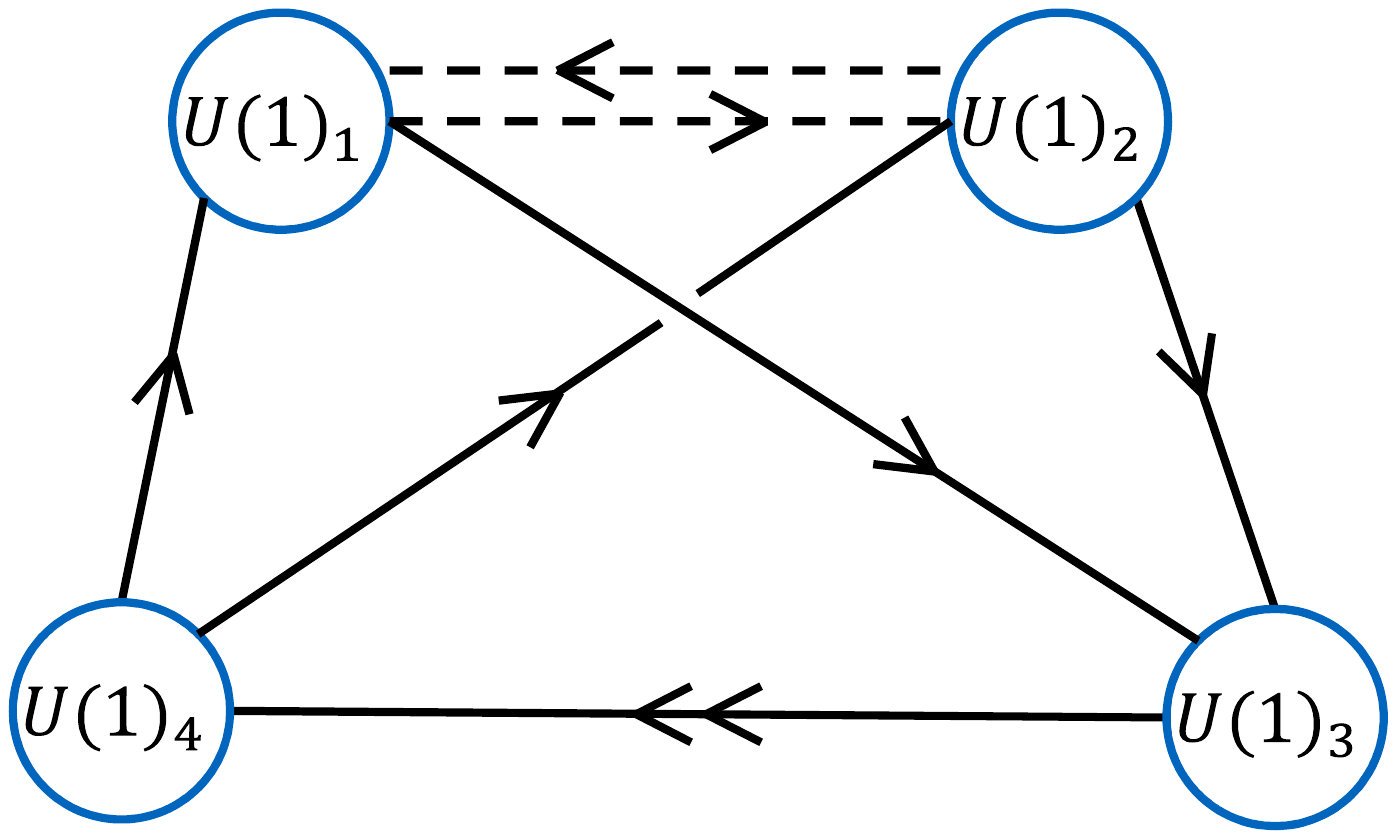}
\caption{
The left panel: $U(2) \times U(1)_{1} \times U(1)_{2}$ quiver diagram.
The right panel: $U(1)_{1} \times U(1)_{2} \times U(1)_{3} \times U(1)_{4}$ quiver diagram 
obtained from $U(2) \to U(1)_{1} \times U(1)_{2}$ by the Higgs mechanism.
The dashed quiver shows bi-fundamental scalars arising from this symmetry breaking.
}
\label{fig:quiverN34ex1}
\end{figure}
%
\begin{figure}[t]
\centering
\includegraphics[scale=0.3]{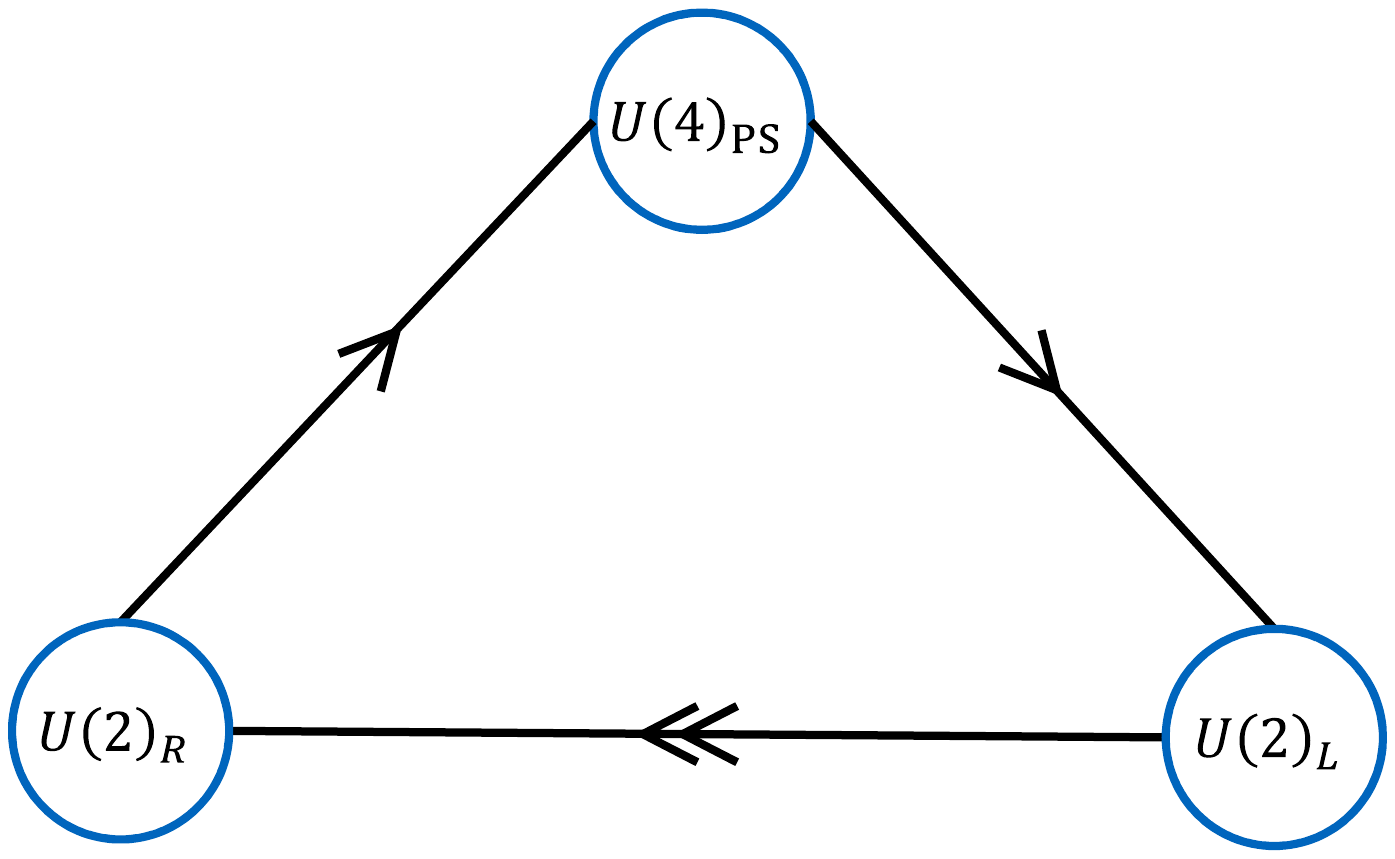}
\caption{
A quiver diagram of left-right symmetric Pati-Salam model.
}
\label{fig:Pati-Salam}
\end{figure}
%

%
We shall consider the $U(1)^4$ gauge theory shown in the right panel of Fig.~\ref{fig:quiverN34ex1}.
It is noted that we have two left-handed fermions charged only under $U(1)_3 \times U(1)_4$,
and there exist six chiral fermions and two complex scalars. 
This model is obtained from three nodes model of 
$U(2) \times U(1)_{3} \times U(1)_{4}$ in the left panel of Fig.~\ref{fig:quiverN34ex1} 
by the Higgs mechanism of $U(2)$ complex adiont scalar whose vacuum expectation value is given by 
$\braket{\Phi} = \mathrm{diag} (v , -v)$. 
This can be regard as a toy model of left-right symmetric theory obtained 
from the Pati-Salam model \cite{Pati:1974yy,Mohapatra:1974hk,Senjanovic:1975rk} as in Fig.~\ref{fig:Pati-Salam}.
The $U(1)^4$ model has two anomaly-free $U(1)$'s and non-trivial charged matter fields,
but we focus only on the relevant gauge couplings.
The detail of the anomaly cancellation is discussed in Appendix~\ref{appA}.
The divergences of $U(1)$ currents are given by
\begin{align}
 \partial \cdot
 \left( \begin{array}{c}
 j^{1} \\
 j^{2} \\
 j^{3} \\
 j^{4}
 \end{array}\right)
 = 
 \left( \begin{array}{cccc}
 0 & 0 & 1 & -1 \\
 0 & 0 & 1 & -1 \\
 -1 & -1 & 0 & 2 \\
 1 & 1 & -2 & 0
 \end{array}\right)
 \left( \begin{array}{c}
 Q_{1} \\
 Q_{2} \\
 Q_{3} \\
 Q_{4}
 \end{array}\right),
\end{align}
and we define two independent anomaly-free $U(1)$'s with a free parameter $c$ as
\begin{align}
 U(1)_{X} = & c U(1)_{1} + (2- c) U(1)_{2} +U(1)_{3} +U(1)_{4},
 \\
 U(1)_{X'} = & \frac{3-c}{1-c} U(1)_{1} - \frac{1+c}{1-c} U(1)_{2} + U(1)_{3} + U(1)_{4}.
\end{align}
Two chiral fermions charged only under $U(1)_3 \times U(1)_4$ is still neutral
but other fermions have non-trivial charges under these anomaly-free $U(1)$ gauge groups. 
The corresponding anomaly-free gauge couplings read
\begin{align}
 \frac{1}{e_{X}^2} = & \frac{c^2}{g_{1}^2} + \frac{(2-c)^2}{g_{2}^2} + \frac{1}{g_{3}^2} + \frac{1}{g_{4}^2},
 \\
 \frac{1}{e_{X'}^2} =& \biggl( \frac{3-c}{1-c} \biggr)^2 \frac{1}{g_{1}^2} + \biggl( \frac{1+c}{1-c} \biggr)^2 \frac{1}{g_{2}^2}
 + \frac{1}{g_{3}^2} + \frac{1}{g_{4}^2}.
\end{align}
These gauge couplings are given by linear combinations of the original ones,
and when $U(1)_1 \times U(1)_2$ is unified to $U(2)$ we find $g_1 = g_2$.
\subsection{A model inspired by the SM}
%
\begin{figure}[t]
\centering
\includegraphics[scale=0.26]{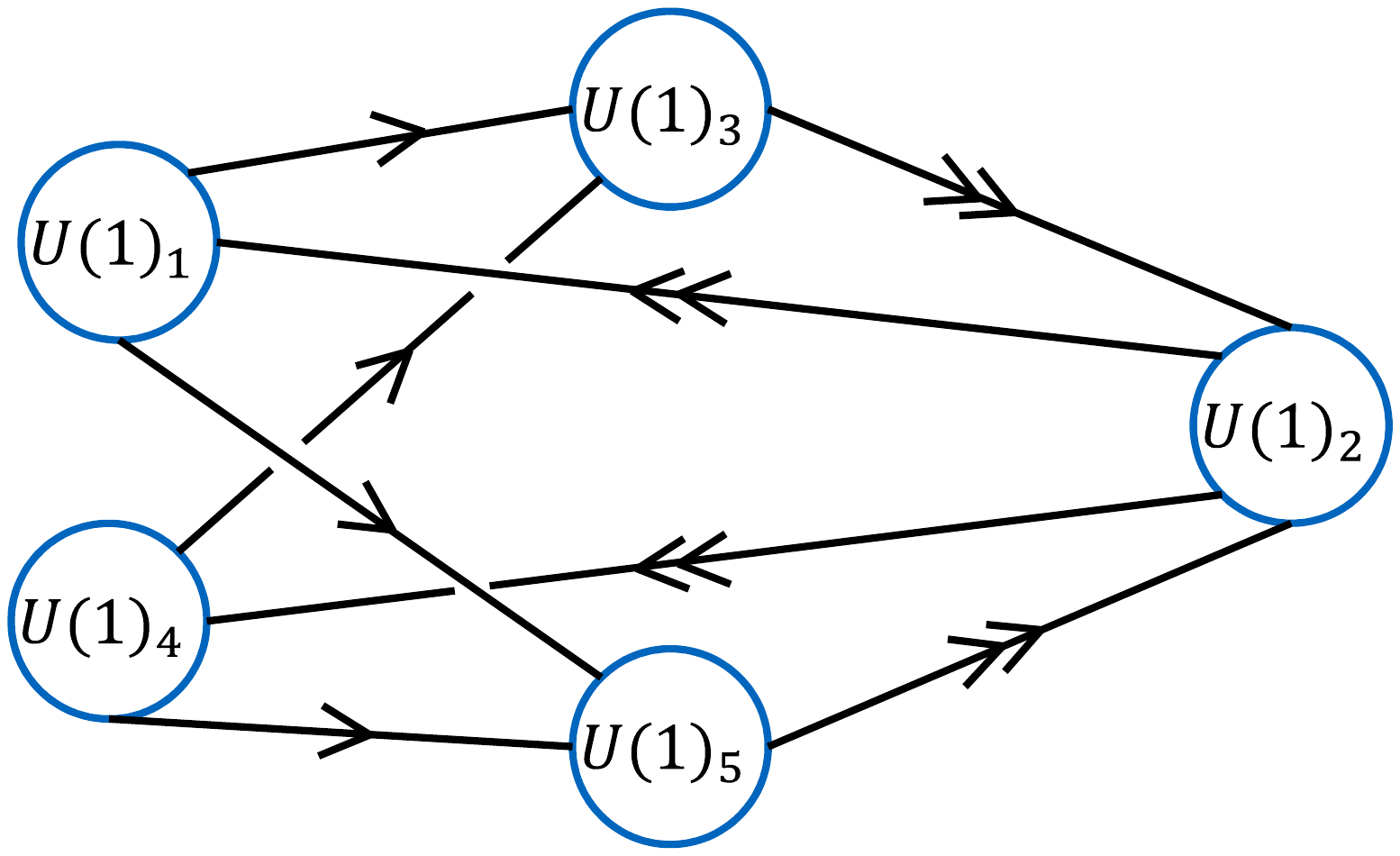}
\caption{
A diagram of quiver gauge theory inspired by the SM.
}
\label{fig:quiverN5ex1}
\end{figure}
%
\begin{figure}[t]
\centering
\includegraphics[scale=0.29]{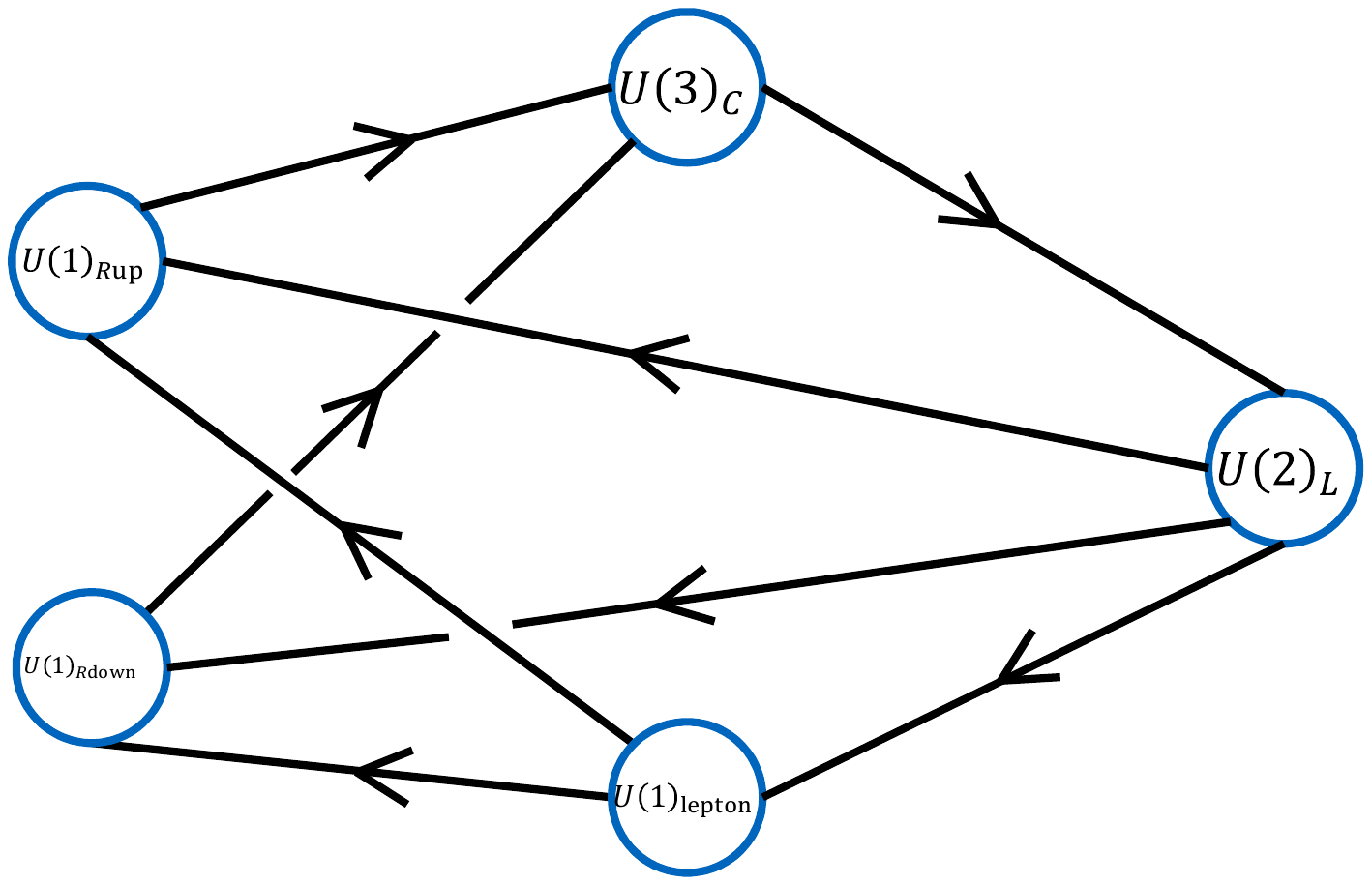}
\caption{
A quiver diagram of the SM-like model \cite{Verlinde:2005jr}.
}
\label{fig:quiverSM}
\end{figure}
%

%
Another example is a model in Fig.~\ref{fig:quiverN5ex1} that is
inspired by the SM-like model in Fig.~\ref{fig:quiverSM}\footnote{
The authors of Ref.~\cite{Verlinde:2005jr} discussed 
the world-volume theory on a stack of D3-branes reproducing the field content 
of the minimal supersymmetric standard model with extended Higgs sector in a quiver extension. 
}.
The divergences of chiral currents are given by
\begin{align}
 \partial \cdot \left( \begin{array}{c}
 j^{1} \\
 j^{2} \\
 j^{3} \\
 j^{4} \\
 j^{5}
 \end{array}\right)
 = 
 \left( \begin{array}{ccccc}
 0 & -2 & 1 & 0 & 1 \\
 2 & 0 & -2 & 2 & -2 \\
 -1 & 2 & 0 & -1 & 0 \\
 0 & -2 & 1 & 0 &1 \\
 -1 & 2 & 0 & -1 & 0
 \end{array}\right)
 \left( \begin{array}{c}
 Q_{1} \\
 Q_{2} \\
 Q_{3} \\
 Q_{4} \\
 Q_{5}
 \end{array}\right).
 \label{eq:quiverN5ex1}
\end{align}
We find three anomaly-free $U(1)$'s and they can generally be written as
\begin{align}
 U(1)_{X} = c_{1} U(1)_{1} + U(1)_{2} + c_{2} U(1)_{3} + (2 -c_{1}) U(1)_{4} + (2-c_{2}) U(1)_{5},
\end{align}
with two free parameters of $c_1$ and $c_2$ which will be a rational numbers.
The parameters of $c_i$'s are taken as a gauge group is orthognal to each other.
Then the fermions have non-trivial charge in this anomaly-free $U(1)$'s, but we focus only on
the gauge couplings. 
The relavant gauge coupling is given by
\begin{align}
 \frac{1}{e_{X}^2} = \frac{c_{1}^2}{g_{1}^2} + \frac{1}{g_{2}^2} + \frac{c_{2}^2}{g_{3}^2} + \frac{(2-c_{1})^2}{g_{4}^2} 
+ \frac{(2-c_{2})^2}{g_{5}^2}.
\end{align}
As mentioned earlier, an anomaly-free coupling can contain more of the original couplings
as the number of $U(1)$'s in a theory increases.
%

\section{Orbifold compactification}
\label{appC}

Let us dimensionally reduce the 5D action in Eq.~(\ref{5DS}) and
show the gauge couplings and Yukawa couplings in 4D.
The metric of $M_4 \times S^1 / \mathbb{Z}_2$ is written by 
$ds_5^2 = e^{-\sigma} g_{\mu\nu} dx^\mu dx^\nu + e^{2 \sigma}dy^2$.
Thus, the vielbein is given as
\begin{align}
 E^{A}{}_{M} = \left( \begin{array}{cc}
 e^{ - \sigma /2} e^{a}{}_{\mu} & \\
 & e^{\sigma}
 \end{array}\right),
\end{align}
where $A$ and $a$ represent 5D and 4D local Lorentz indices respectively,
and $e^a{}_\mu$ is the 4D vielbein. 
The off-diagonal element of the vielbein is absent because the orbifold projection prohibits the graviphoton.
It is noted that 5D fermion kinetic term is given by 
$\bar\Psi \Gamma^A E_{A}{}^{M} D_M \Psi$, where $E_{A}{}^{M}$ is the inverse matrix of $E^{A}{}_{M}$.
Using these equations,
we obtain 4D action for massless modes in Eqs.~(\ref{eq:orbifoldA}) and (\ref{eq:orbifoldpsi}):
\begin{align}
 &S_{4 \mathrm{D}}
 = \int d^4x\,\sqrt{-g_{4}} \Biggl[
 \frac{\pi L}{2\kappa_{5}^2} \mathcal{R}_{4} - \frac{3\pi L}{4 \kappa_{5}^2} (\partial_{\mu} \sigma)^2
 \nonumber\\
 &
 - \frac{1}{4}\frac{\pi L e^{\sigma}}{\hat{g}_a^2}(F^{(a)}_{\mu\nu})^2 -2\frac{1}{4} \frac{\pi L e^{\sigma}}{\hat{g}_2^2} (F^{(b)}_{\mu\nu})^2 -\frac{1}{4} \frac{\pi L e^{\sigma}}{\hat{g}_{c}^2} (F^{(c)}_{\mu\nu})^2 - 2\frac{1}{4} \frac{ \pi L e^{\sigma}}{\hat{g}_{2}^2} (F^{(d)}_{\mu\nu})^2
 - \frac{ \pi L e^{-2\sigma}}{ \hat{g}_2^2} |D_\mu \varphi|^2
 \nonumber\\
 &
 + \pi L e^{-\sigma/2} i \bar{\psi_{ab}} \Slash{D} \psi_{ab}
 + \pi L e^{-\sigma/2} i \bar{\psi_{da}} \Slash{D} \psi_{da}
 +\pi Le^{-\sigma/2} i \bar{\psi_{cd}} \Slash{D} \psi_{cd}
 + \pi L e^{-\sigma/2} i \bar{\psi_{bc}} \Slash{D} \psi_{bc}
 \nonumber\\
  &
 -\frac{\pi L e^{-2\sigma}}{\sqrt{2}} \varphi_{R} (\bar{\psi_{ab}^{C}} \psi_{da} +\bar{\psi_{da}} \psi_{ab}^{C})
 - \frac{\pi L i e^{-2\sigma}}{\sqrt{2}} \varphi_{I} ( \bar{\psi_{ab}^{C}} \psi_{da} - \bar{\psi_{da}} \psi_{ab}^{C})
 \nonumber\\
 &
 -\frac{\pi L e^{-2\sigma}}{\sqrt{2}} \varphi_{R} ( \bar{\psi_{bc}^{C}}\psi_{cd} +\bar{\psi_{cd}} \psi_{bc}^{C})
 + \frac{\pi L i e^{-2\sigma}}{\sqrt{2}} \varphi_{I} ( \bar{\psi_{bc}^{C}} \psi_{cd} - \bar{\psi_{cd}} \psi_{bc}^{C} )
 -V ( \sigma, \varphi)
 \Biggr],
\end{align}
where 
$D_{\mu} = \partial_{\mu} +i \sum_{j= a,b,c,d} q_{\psi} A^{(j)}_{\mu}$ 
is the covariant derivative associated with the gauge group $U(1)_{a} \times U(1)_{b} \times U(1)_{c} \times U(1)_{d}$,
and the chiral fermions $\psi_{ij}$ are defined in Sec.~\ref{sec:orbifoldmodel}.
The four dimensional Ricci scalar is denoted by $\mathcal{R}_{4}$, and
$\varphi = \varphi_R + i \varphi_I$ is the complex scalar.
We introduce the scalar potential $V(\sigma, \varphi)$ formally\footnote{
At the classical level, the scalars $\sigma$ and $\varphi$ do not have potential due to the gauge symmetries, 
but the potential can be generated by the radiative corrections.
In addition, we assume the radion field develops the VEV of $\braket{\sigma} =0$ around the radius $L$.
}.
Thus, the gauge couplings are given as in Eq.~(\ref{eq:gaugecoupling}).
In order to find the relation between the Yukawa coupling and the gauge coupling, 
we canonically normalize the fermion and the complex scalar as
\begin{align}
 \psi_{ij} \to \frac{ e^{\sigma/4}}{\sqrt{\pi L}} \psi_{ij},
 \quad
 \varphi \to \frac{ \hat{g}_{2} e^{\sigma}}{\sqrt{\pi L}} \varphi.
 \label{eq:rescaling}
\end{align}
Then, the kinetic term is rewritten as
\begin{align}
 \pi L e^{-\sigma /2} i \bar{\psi_{ij}} \Slash{D} \psi_{ij}
 \to i \bar{\psi_{ij}} \Slash{D} \psi_{ij} 
 - \frac{i}{2} \bar{\psi_{ij}} \gamma^\mu \psi_{ij} \partial_\mu \sigma .
\end{align}
Hereafter, we will ignore the derivative coupling of the radion to the fermions.
The Yukawa interactions are expressed as
\begin{align}
 \mathcal{L}_{\mathrm{4D,Yukawa}} = &
 -\frac{ \hat{g}_{2} e^{-\sigma/2}}{\sqrt{2 \pi L }} \varphi_{R} (\bar{\psi_{ab}^{C}} \psi_{da} +\bar{\psi_{da}} \psi_{ab}^{C})
 - \frac{\hat{g}_{2} i e^{-\sigma/2}}{\sqrt{2 \pi L}} \varphi_{I} ( \bar{\psi_{ab}^{C}} \psi_{da} - \bar{\psi_{da}} \psi_{ab}^{C})
 \nonumber\\
 &
 -\frac{\hat{g}_{2} e^{-\sigma/2}}{\sqrt{2 \pi L }} \varphi_{R} ( \bar{\psi_{bc}^{C}}\psi_{cd} +\bar{\psi_{cd}} \psi_{bc}^{C})
 + \frac{\hat{g}_{2} i e^{-\sigma/2}}{\sqrt{2 \pi L}} \varphi_{I} ( \bar{\psi_{bc}^{C}} \psi_{cd} - \bar{\psi_{cd}} \psi_{bc}^{C} ).
  \nonumber\\
\end{align}
We introduce the Dirac fermions as
\begin{align}
 \psi_{a} = \left( \begin{array}{c}
 \psi_{da} \\
 \psi_{ab}^{C}
 \end{array}\right).
 \quad
 \psi_{c} = \left( \begin{array}{c}
 \psi_{cd} \\
 \psi_{bc}^{C}
 \end{array}\right).
\end{align}
With these Dirac fermions, 
the kinetic terms of the anomaly-free sector read
\begin{align}
 \mathcal{L}_{\mathrm{4D,KT}} = &
 - \frac{1}{4 e_{X}^2} \bigl( F^{(X)}_{\mu\nu} \bigr)^2 -\frac{1}{4 e_{X'}^2} \bigl( F^{(X')}_{\mu\nu} \bigr)^2
 - (\partial_\mu \varphi_{R})^2 - ( \partial_{\mu} \varphi_{I} )^2
 \nonumber\\
 & + i \bar{\psi_{a}} \Slash{\mathcal{D}} \psi_{a} + i \bar{\psi_{c}} \Slash{\mathcal{D}} \psi_{c},
 \label{eq:kineticterm}
\end{align}
where we neglected gauge bosons in anomalous $U(1)$'s,
the anomaly-free gauge couplings are defined by Eqs.~(\ref{eq:g_{X}^2SWGC}) and (\ref{eq:g_{X'}^2SWGC}).
As shown in Table~\ref{tab:quiver_SWGC},
the covariant derivatives of the Dirac fermions associated with anomaly-free $U(1)$'s are expressed as
\begin{align}
 \mathcal{D}_{\mu} \psi_{a} = & \biggl[ \partial_{\mu} + i (-1 +c ) A^{(X)}_{\mu} - i \biggl( 1 + \frac{1}{c} \biggr) A^{(X')}_{\mu} \biggr] \psi_{a},
 \\
 \mathcal{D}_{\mu} \psi_{c} = & \biggl[ \partial_{\mu} - i ( -1 + c) A^{(X)}_{\mu} + i \biggl( 1 + \frac{1}{c} \biggr) A^{(X')}_{\mu} \biggr] \psi_{c}.
\end{align}
The charges of $\psi_{a}$ and $\psi_{c}$ under $U(1)_{X}$ and $U(1)_{X'}$ are opposite
to each other
due to the 4D anomaly-free conditions.
With the Dirac spinor, the Yukawa terms in this Lagrangian are rewritten as below:
\begin{align}
 \mathcal{L}_{\text{4D,Yukawa}} =&
 - y e^{-\sigma/2} \varphi_{R} \bar{\psi_{a}} \psi_{a}
 - i y e^{ - \sigma/2} \varphi_{I} \bar{\psi_{a}} \gamma_{5} \psi_{a}
 - y e^{-\sigma /2} \varphi_{R} \bar{\psi_{c}} \psi_{c}
 + i y e^{- \sigma/2} \varphi_{I} \bar{\psi_{c}} \gamma_{5} \psi_{c},
\end{align}
where the 4D Yukawa coupling is defined as
\begin{align}
 y = \frac{\hat{g}_{2}}{\sqrt{ 2\pi L}}.
\end{align}

\end{document}